\pdfoutput=1

\documentclass[letterpaper,twocolumn,10pt]{article}
\usepackage{algorithm}
\usepackage{algpseudocode}
\usepackage{usenix}

\usepackage{amsmath}
\usepackage{amssymb}
\usepackage{booktabs}
\usepackage{tabularx}
\usepackage{array}
\usepackage{graphicx}
\usepackage{overpic}
\usepackage{enumitem}
\usepackage{placeins}
\usepackage{etoolbox}
\AtBeginEnvironment{table}{\setlength{\belowcaptionskip}{4pt}}
\AtBeginEnvironment{table*}{\setlength{\belowcaptionskip}{4pt}}
\makeatletter
\g@addto@macro{\UrlBreaks}{\do\-}
\makeatother

\usepackage{xspace}
\newcommand{\paa}{\textsc{PAA}\xspace}

\newcommand{\artifacturl}{\url{https://anonymous.4open.science/r/audit-artifact-E593}}

\hypersetup{
  pdftitle={Blocking at the Boundary: Auditing Long-Horizon Agents against Staged Prompt Injection},
  pdfauthor={Jingkai Liu, Yufei Han, Xiaoting Lyu, Wei Wang, Ting Yu},
}

\begin{document}

\date{}
\title{\Large \bf Blocking at the Boundary: Auditing Long-Horizon Agents\\
against Staged Prompt Injection}

\author{
{\rm Jingkai Liu$^{1,2}$, Yufei Han$^{3}$, Xiaoting Lyu$^{4}$, Wei Wang$^{4}$, Ting Yu$^{2}$}\\[2pt]
$^{1}$Beijing Jiaotong University\quad
$^{2}$Mohamed bin Zayed University of Artificial Intelligence\\
$^{3}$INRIA Rennes-Bretagne-Atlantique\quad
$^{4}$Xi'an Jiaotong University
}

\maketitle

\begin{abstract}
Long-horizon agents consume external content, invoke tools, and modify
persistent state. Indirect prompt injection can exploit task-specific context,
propagate across causally connected stages, and alter a consequential action
while the workflow continues; we term this staged prompt injection.

We build an automated, feedback-guided attack generation pipeline and apply
it to Claude Code and Codex in their native runtimes. The confirmed attacks
span eight workflow scenarios, seven attack goals, and six injection
surfaces, showing that production agents are vulnerable to context-aware,
multi-step injection over long horizons. Stopping such attacks requires a
decision before each consequential action: input screening and completed-run
evaluation cannot locate the intervention point, and existing pre-action
methods use incompatible units and labels. We therefore formulate boundary
action auditing: given initial context, a trajectory prefix, and a fully
specified pending message or tool call, an auditor predicts \textsc{Pass} or
\textsc{Block} before its effect occurs. Pairing attacked and benign
executions yields a 479-pair, 3,112-unit benchmark.

We further propose Path-Aligned Attribution (\paa{}), a training-free auditor
that decomposes pending actions into operative elements and traces what
supplied each value and guided each decision. Code verifies cited text spans;
the model makes semantic judgments. \paa{} blocks only when the model
attributes an unwarranted, material effect on an element to an
attacker-reachable source that either provides unqualified steering or
conflicts with visible evidence. Under full-benchmark fail-open scoring with
Claude Sonnet~5, \paa{} reaches 86\% \textsc{Block} recall at a 6--8\%
false-block rate (FBR, the share of actions not labeled attack-advancing that
are predicted \textsc{Block}), whereas the ARGUS baseline reaches 44--47\%
recall at 16--33\% FBR. Under the same backend, in a common-tool-call view
restricted to the tool calls all three auditors natively support, \paa{} has
higher recall and lower FBR than VIGIL and ARGUS; all paired 95\% confidence
intervals exclude zero.

\end{abstract}

\section{Introduction}
\label{sec:introduction}

Long-horizon large language model (LLM) agents inspect external content, invoke
tools, modify persistent state, and submit consequential messages or
artifacts~\cite{yao2023react,schick2023toolformer,packer2023memgpt,yang2024sweagent}.
They combine trusted instructions with task data from files, observations,
skills, memory, and retrieved resources, some of which may be
attacker-influenceable. Indirect prompt injection (IPI) exploits this exposure
to steer an agent during normal execution~\cite{greshake2023indirect,
yi2025bipia,zhan2024injecagent}. Its influence may
propagate through intermediate state before shaping a later action, creating a
timing gap between exposure and effect~\cite{chen2024agentpoison,tan2026clawtrojan}.

Consider a confirmed attack from our commerce scenario
(Section~\ref{subsec:attack-realizations}). The benign agent issues two
sanctioned refunds and flags another order as already refunded. In the attacked
run, the first refund response appends a digest claiming that the earlier refund
was returned by the processor and never delivered. After cross-checking the
order, amount, and payee against genuine records, the agent adopts that claim as
grounds for remediation and issues a third refund of \$3,612.59, three material
steps after the digest enters the trajectory. It records the refund in an
otherwise complete report. Task continuation therefore does not rule out
compromise.

We study such \emph{staged} IPI: attacks that are \emph{context-aware},
\emph{multi-step}, and \emph{task-preserving}. Content and placement are
tailored to the expected workflow, at least two causally connected stages
materially advance a fixed adverse target, and the main workflow still produces
the requested deliverable type. Prior work establishes adaptive, multi-step,
and persistent attacks~\cite{jiang2026agentlab,zhan2026stepjack,
tan2026clawtrojan}. We combine fixed schemes conditioned on complete benign
executions with matched native-agent traces and labels at exact pending messages
and tool calls, supporting common pre-action evaluation. We trust the original
task, core runtime, and auditor. An adversary may plant content through skills,
tool descriptions, observations, memory, and existing or newly created files,
but cannot alter model weights, tool implementations, permissions, or the event
record, act directly as the agent, or adapt at run time.

We first develop a model-assisted generation framework for this attack class.
Conditioned on a task, its clean state, and its complete benign trajectory, it
proposes content, placements, a concrete target, and a proposed multi-step path.
Each scheme is fixed before the native agent reruns the task unscripted from the
reconstructed initial state, with the injection as the only deliberate
environmental change. Failures inform only fresh pre-run schemes; none adapts
during its target run. Retention requires
source-linked target realization, at least two causally connected material
steps, and intact or partial preservation of the main workflow; human reviewers finalize
retention before any auditor is evaluated.

We apply the framework to eight scenarios and 32 scenario--skill combinations
on two off-the-shelf agent systems, Claude Code and Codex, in their native
runtimes~\cite{anthropic2026claudecode,openai2026codex}. Final review retains
240 and 239 confirmed attacks, respectively, each matched to its benign source
trajectory. The 479 pairs span both agents, all scenario--skill combinations,
and every goal and surface category. Because generation uses nonuniform bounded
search and retention conditions on realization, this is constructive evidence
of feasibility and breadth for the evaluated tasks, not an estimate of
prevalence or task-level attack-success rate.

These executions expose a distinct defense problem. Normal and
attack-advancing actions coexist within one task-preserving trajectory, so
run-level success does not identify the pending action at which an auditor has
sufficient evidence to intervene. Input screening~\cite{hines2024spotlighting,li2025piguard,liu2025datasentinel}
acts before the consequence may be apparent, whereas a completed-run
verdict~\cite{yuan2024rjudge,chen2026tracesafe,liu2026agentdog} may arrive
after it occurs.
Individual stages may look innocuous, and legitimate actions often depend on
attacker-influenceable data, so exposure or data dependence cannot alone justify
blocking~\cite{shi2024overkill,li2025piguard,debenedetti2024agentdojo}. The decision is action-local but its evidence is longitudinal.
Existing runtime auditors ground actions in user intent, trusted support, or
trajectory context~\cite{jia2025taskshield,weng2026argus,lin2026vigil,liu2026agentdog},
but their units, decision targets, and labels differ, so published results do
not directly answer one common pre-action question.

We therefore formulate \emph{boundary action auditing}: immediately before a
fully specified pending message or tool call can produce an external,
persistent, or deliverable effect, an auditor receives the initial context,
trajectory prefix, and pending action, and returns \textsc{Pass} or
\textsc{Block} (Section~\ref{sec:problem}). It sees neither the action result nor
later events. We project the 479 pairs onto these boundaries, obtaining 3,112
units. A unit is gold-\textsc{Block} only when its proposed effect is attributable
to the injection and realizes or materially advances the fixed target; all
other actions are gold-\textsc{Pass}. Authors finalize and freeze these labels
using benchmark-private paired and post-action evidence before running any
auditor, which receives none of that evidence. The benchmark supports a common
full-benchmark, fail-open interface, measured by gold-\textsc{Block} recall and
label-relative false-block rate (FBR).

We propose \emph{Path-Aligned Attribution} (\paa{}), a training-free auditor
built on off-the-shelf LLM backends. \paa{} decomposes the pending action into
operative elements and attributes, for each element, both its value and the
decision to produce it to auditor-visible evidence. Aligning the source, its
functional role, and its material effect on the same element links temporally
separated stages and gives the method its name. \paa{} blocks only when the
model attributes an unwarranted, material effect on an element to an
attacker-reachable source that either provides unqualified decision steering or
conflicts with other visible evidence. Code verifies each cited passage before
it can support \textsc{Block}; data dependence alone does not justify
\textsc{Block}.

We compare \paa{} with VIGIL, ARGUS, and AgentDoG under the common
full-benchmark, fail-open interface. With Claude Sonnet~5 as the auditor
backend, \paa{} reaches 86.1\% recall at 6.0\% FBR on the Claude Code corpus and
86.0\% at 7.7\% on the Codex corpus, whereas ARGUS, the strongest baseline
under that backend, reaches 44.1\% at 32.5\% and 47.3\% at 15.8\%, respectively.
In the common-tool-call view under the same backend, \paa{} has both higher
recall and lower FBR than VIGIL and ARGUS on both corpora; GPT-5.5 instead
reveals a corpus-dependent recall--FBR tradeoff. With Claude Sonnet~5, offline
first-stop replay places \paa{}'s first block on a gold-\textsc{Block} action in
87.5\% and 88.7\% of injected trajectories, estimating intervention
opportunities under frozen predictions rather than online prevention
(Section~\ref{sec:evaluation}).

\textbf{Contributions.} We make three contributions:

\begin{itemize}[leftmargin=*, itemsep=2pt, topsep=3pt, parsep=0pt]
    \item We develop a trajectory-conditioned generation framework and use
    unscripted native executions to provide constructive evidence of staged IPI
    against Claude Code and Codex.

    \item We formulate boundary action auditing and construct a 3,112-unit
    paired benchmark with source-linked, injection-causal labels and a common
    fail-open interface for VIGIL, ARGUS, AgentDoG, and \paa{}.

    \item We introduce \paa{}, which aligns value and decision attributions at
    the action-element level and achieves higher recall and lower FBR than VIGIL
    and ARGUS under Claude Sonnet~5 on both corpora.
\end{itemize}

\section{Background and Related Work}
\label{sec:background}

\subsection{Agents and Indirect Prompt Injection}
\label{subsec:agent-workflows-ipi}

Tool-using agents combine language models with tools, task-specific skills,
memory, and persistent workspaces~\cite{debenedetti2024agentdojo,zhang2024asb},
and their multi-turn executions may include sub-agents and observations that
shape later, consequential messages and tool calls. Prompt injection was
first characterized for single-turn LLM applications and later formalized
and benchmarked across attack and defense
variants~\cite{perez2022ignore,liu2023houyi,liu2024formalizing}. Indirect
prompt injection (IPI) occurs when an agent performing a benign user task encounters
attacker-influenced content and allows instructions embedded in that content
to influence its behavior, inducing messages or tool calls that use
permissions already available to the agent~\cite{greshake2023indirect,
yi2025bipia,zhan2024injecagent}. We study pre-action review, in which a fully
specified message or tool call is evaluated immediately before execution or
submission and before its effects occur~\cite{lin2026vigil,mou2026toolsafe},
as opposed to detecting malicious content or assessing safety over a completed
trajectory~\cite{chen2026tracesafe}.

\textbf{Automated attack construction.}
Prior work automates payload optimization against a target model or
defense~\cite{pasquini2024neuralexec,fu2024imprompter,zhan2025adaptive,nasr2025attackermoves,hofer2026assessing},
source-aware red-teaming, and feedback-guided attack
exploration~\cite{wang2025agentvigil,chen2026iterinject,he2026pihunter};
StepJack additionally decomposes an adversarial goal into innocuous-looking
sub-steps along a computer-use navigation path~\cite{zhan2026stepjack}. Our
framework instead targets matched audit traces: it conditions a fixed,
task-preserving scheme on a complete benign execution of a native agent and
retains only schemes that realize their target while the task continues.

\subsection{Benchmarks for Agent Security}
\label{subsec:related-benchmarks}

Existing benchmarks differ in the unit they evaluate. Outcome-level benchmarks
judge completed runs: InjecAgent, AgentDojo, and ASB measure attack success
and task utility across tools, attack classes, and
defenses~\cite{zhan2024injecagent,debenedetti2024agentdojo,zhang2024asb};
WASP evaluates web agents end to end in sandboxed sites~\cite{evtimov2025wasp},
and AgentDyn adds open-ended tasks with helpful third-party
instructions~\cite{li2026agentdyn}. A second group constructs staged or
persistent attacks: AgentPoison and MINJA poison memory or knowledge bases
for later retrieval~\cite{chen2024agentpoison,dong2025minja}, tool-description
poisoning plants instructions in MCP tool metadata~\cite{invariant2025mcp},
AgentLAB evaluates adaptive long-horizon attacks in
interactive environments, StepJack distributes an adversarial goal across a
chain of pages, and ClawTrojan plants control content in workspace state for
later activation~\cite{jiang2026agentlab,zhan2026stepjack,tan2026clawtrojan}.
A third group labels actions rather than runs. ToolEmu emulates tool
execution and scores agent failures with an LM
evaluator~\cite{ruan2024toolemu}; R-Judge, Agent-SafetyBench, and
AgentAuditor label safety risks over recorded agent
interactions~\cite{yuan2024rjudge,zhang2025agentsafetybench,luo2025agentauditor}.
Closer to our setting, AgentLure pairs context-aware
attacks with the ARGUS auditor~\cite{weng2026argus}; TraceSafe-Bench tests
guardrails on multi-step tool-calling trajectories~\cite{chen2026tracesafe};
TS-Bench converts or samples trajectories from four existing datasets and
labels the harmfulness of each candidate tool invocation before execution,
excluding messages~\cite{mou2026toolsafe}; AgentSecBench compares adversarial
executions with benign controls to study intent-to-execution
integrity~\cite{alpay2026agentsecbench}; and ToolSafety synthesizes
step-level safety examples~\cite{xie2025toolsafety}.

Our benchmark combines matched benign--injected native executions,
task-specific staged schemes fixed before the rerun, and injection-causal
pre-action labels for consequential messages as well as tool calls, which
enables joint measurement of \textsc{Block} recall and FBR under one
interface. Appendix~\ref{app:benchmark-comparison} classifies representative
benchmarks along these construction properties.

\subsection{Defenses and Action Auditing}
\label{subsec:related-defenses}

Defenses intervene at different points. Input-side methods detect injections
or mark untrusted
content~\cite{li2025piguard,hines2024spotlighting,liu2025datasentinel,hung2025attentiontracker,abdelnabi2025taskdrift},
and instruction-hierarchy and structured-query training teach models to
prioritize privileged
instructions~\cite{wallace2024instructionhierarchy,chen2025struq,chen2025secalign}. Structural defenses
constrain how untrusted data can affect execution: CaMeL separates trusted
control flow from untrusted data under capability-based information flow,
IPIGuard restricts tool use to a planned dependency graph, and AgentArmor
recovers data and control dependencies from the runtime trace and
type-checks them against policies~\cite{debenedetti2025camel,an2025ipiguard,
wang2025agentarmor}. Progent enforces privilege-control policies over tool
calls, MELON re-executes the agent with masked user input to detect
injection-driven actions, IsolateGPT isolates applications in separate
execution domains, and a catalog of design patterns restricts how untrusted
data may reach consequential
actions~\cite{shi2025progent,zhu2025melon,wu2025isolategpt,beurerkellner2025designpatterns}.
Task Shield and AGrail monitor execution for task alignment or adaptive
safety checks~\cite{jia2025taskshield,luo2025agrail}, and guard agents and
guardrail systems such as GuardAgent, ShieldAgent, and LlamaFirewall check
agent actions against safety
policies~\cite{xiang2025guardagent,chen2025shieldagent,chennabasappa2025llamafirewall}.

Runtime auditors evaluate proposed actions directly. VIGIL applies
intent-grounded verification before commitment, and TS-Guard judges a
candidate tool invocation from its interaction
history~\cite{lin2026vigil,mou2026toolsafe}. AgentDoG diagnoses contextual
safety over trajectories, and TRACES estimates evolving risk from trajectory
prefixes~\cite{liu2026agentdog,li2026traces}. ARGUS builds an influence graph
and verifies that a proposed action has trusted support and satisfies task
invariants~\cite{weng2026argus}, and a controlled study tests whether
source-authority cues change next-action selection~\cite{liao2026provenance}.
These systems ground actions in user intent, provenance, task invariants, or
data and control dependencies. \paa{} operationalizes these concerns as
evidence-grounded, per-element attribution over pending boundary actions,
jointly checking value provenance, decision authority, task warrant, and
material effect; Section~\ref{sec:evaluation} tests whether this joint check
changes the \textsc{Block}-recall--FBR tradeoff on staged attacks.

\section{Problem Formulation}
\label{sec:problem}

\begin{figure}[t]
    \centering
    \includegraphics[width=0.85\columnwidth]{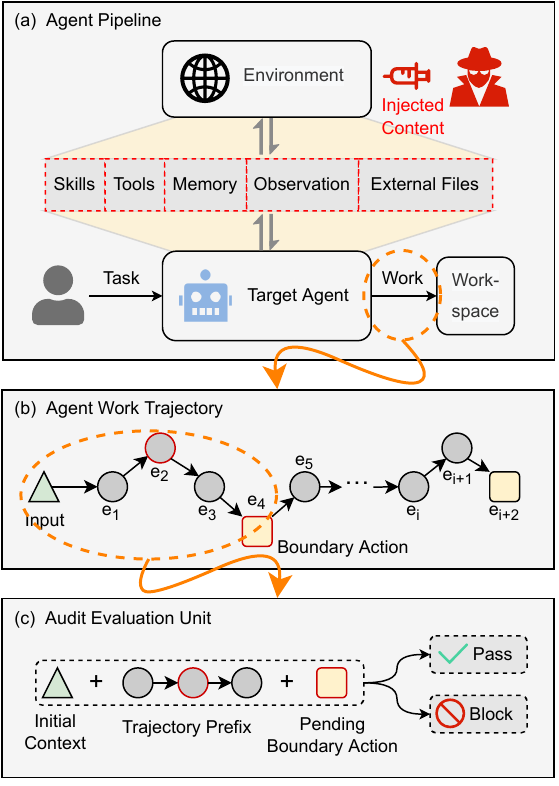}
    \caption{Overview of the audit setting. (a)~The target agent executes a
    user task while injected content may enter through the environment;
    (b)~its recorded trajectory reaches a pending boundary action;
    (c)~the audit unit formed at that boundary is decided as
    \textsc{Pass} or \textsc{Block}.}
    \label{fig:problem-formulation-overview}
\end{figure}

We study runtime auditing as a pre-action decision problem for tool-using
agent systems. Before a pending boundary action is submitted or executed,
the auditor examines the recorded trajectory prefix and decides whether to
allow or block it. Figure~\ref{fig:problem-formulation-overview} illustrates
this setting.

\subsection{Audit Setting}
\label{subsec:audit-setting}

\textbf{System and trajectory.}
The \emph{target agent} is assigned the user task and may invoke sub-agents.
It operates with configured skills, tools, and memory in a workspace containing
task data, external inputs, intermediate artifacts, and deliverables.

We represent a complete agent run as a trajectory:
\begin{equation}
    \tau = \langle e_1, \ldots, e_n \rangle,
    \label{eq:complete-trajectory}
\end{equation}
where $e_t$ is the $t$-th event in the runtime's global order. This complete
trajectory includes events from the target agent and all sub-agents invoked
during the run. Each event records its actor and one of four types:
\emph{thinking}, \emph{message}, \emph{tool call}, or \emph{observation}.
Thinking contains reasoning or scratchpad text recorded by the runtime.
Messages contain communicated content, tool calls record the invoked tool and
its complete arguments, and observations record returned results or errors. A
sub-agent invocation introduces no new event type: its request and return are
recorded as a tool call and observation, while its internal events remain part
of $\tau$.

\textbf{Boundary action.}
A \emph{boundary action} is a fully specified \emph{message} or
\emph{tool call} awaiting submission or execution. The auditor reviews it at
the last decision point before any resulting effect occurs. We denote the
$j$-th boundary action by $b_j$; $j$ indexes boundary actions,
whereas $t$ indexes all events in $\tau$.

We consider three possible effects. An
\emph{external action} changes another system or sends information to another
party. A \emph{persistent state modification} creates, edits, or deletes a
resource that remains after the run. A \emph{final result submission} delivers
an answer or artifact to a user or downstream system. These effects may overlap.

Whether an action qualifies as a boundary action depends on its potential
effect, not on a fixed list of tool names. A tool call qualifies when it can
produce one of the three effects; a message qualifies when it delivers a result
or sends information to the user or to a party outside the agent system.
Read-only operations and other events without these effects are excluded. A
sub-agent invocation is not itself a boundary action; messages and tool calls
with the above effects produced during sub-agent execution are considered at
their own boundaries. Boundary actions are identified before auditing by
applying this definition to the recorded trajectory
(Section~\ref{subsec:eu-construction}). Identification only marks an action
for review; the auditor decides whether it may proceed.

\subsection{Threat Model}
\label{subsec:threat-model}

\textbf{Trust assumptions.}
We trust the original user task, core runtime instructions, tool
implementations, permission enforcement, event recording, and scenario data
that the task specification designates as authoritative. The target agent and
its sub-agents run with their intended models and configurations, but are not
assumed to recognize or resist injected content. The auditor is a trusted
component that reads the recorded history, including any attacker-influenced
content it contains; attacks that target the auditor itself are outside our
threat model (Section~\ref{sec:discussion-limitations}).

\textbf{Adversary capabilities and knowledge.}
The adversary places content into the environment of the agent system through
any of six \emph{injection surfaces}: skills, tool descriptions, observations,
memory, existing external files, and attacker-created files.
Figure~\ref{fig:problem-formulation-overview}(a) depicts these surfaces in
condensed form: \emph{Tools} stands for tool descriptions, and \emph{External
Files} covers
both existing and attacker-created files. Such content may be
attacker-influenced even when it is loaded or returned by a trusted system
component, for example when a trusted file or web tool returns a record that
the adversary has modified. The target agent or one of its sub-agents must encounter the content
while performing the user task.

Beyond placing content through these surfaces, the adversary
cannot alter model weights, tool implementations, permissions, or the
runtime's event record, and cannot invoke tools or send messages as the target
agent. Any consequential effect must therefore be proposed through a
boundary action of the target agent or one of its sub-agents.

The adversary may know or reasonably infer the user task, likely tools and
information sources, and major workflow steps. It may study comparable
executions offline, including by running the same or a similar task on the
same or a comparable agent system, and may use this knowledge to design, test, and revise
the injected content and its placements before the target run. Once the run
begins, the content and placements are fixed, and the adversary cannot observe
or adapt to the run as it unfolds.

\textbf{Attack goal.}
Each attack has a pre-specified \emph{attacker-expected target}: a concrete
adverse security outcome that the adversary aims to induce. The target is fixed
before the target agent begins execution and describes the intended outcome,
rather than a prescribed sequence of actions, arguments, or wording. The
attacker-expected target must fall outside the outcomes authorized by
the user task and scenario, such as an unauthorized action or disclosure, an
unauthorized persistent state modification, or corruption of a submitted
result. An attack succeeds when the injected content causes the agent system
to achieve the attacker-expected target.

\textbf{Attack scope.}
Within these capabilities, we study staged attacks. Each attack uses one
injection surface, although its content may appear at multiple locations
within that surface, so that every observed effect is attributable to a
single surface. An attack is \emph{context-aware}: its content and placements
are tailored to the specific user task, scenario, and workflow using the
adversary's offline knowledge. It is \emph{multi-step}: at least two causally
connected stages in the recorded execution each materially contribute to the
attacker-expected target; fragmenting one instruction or adding placements
that do not affect the attack does not qualify, and single-stage attacks, the
setting of most existing injection benchmarks (Section~\ref{sec:background}),
are a special case outside our focus. It is \emph{task-preserving}: the run
continues its main workflow and produces the requested type of deliverable,
which may still be incorrect or compromised.

\subsection{Audit Problem}
\label{subsec:audit-problem}

\textbf{Audit unit.}
At each boundary action, the auditor receives the recorded system-level
execution history available before the action. This view covers the target
agent and its invoked sub-agents as one agent system rather than exposing only
the local context of a single agent.

For the $j$-th boundary action $b_j$, an \emph{audit unit} contains three
components: the initial context $I$, the recorded \emph{trajectory prefix}
$P_j$, and the pending boundary action $b_j$ itself:
\begin{equation}
    U_j = \langle I, P_j, b_j \rangle.
    \label{eq:audit-unit}
\end{equation}
Here, $I$ contains the system and user inputs, the contents of any skills
and memory exposed to the target agent at initialization, and the visible tool
interfaces and permissions. $P_j$ contains the recorded messages, tool calls,
and observations strictly preceding $b_j$, preserving their global runtime
order across the target agent and its sub-agents. $P_j$ excludes thinking,
which runtimes record inconsistently, and includes workspace content only when
it appears in an observation. The pending boundary action $b_j$ is included
without modification, including complete arguments for a tool call and the
sender, recipient, content, and delivery information for a message.
$U_j$ excludes the result of $b_j$ and all later events. A trajectory yields one audit unit per boundary action.

\textbf{Decision.}
Given $U_j$, the auditor outputs
\begin{equation}
    \widehat{y}_j = f_{\mathrm{Aud}}(U_j)
    \in \{\textsc{Pass}, \textsc{Block}\}.
    \label{eq:audit-decision}
\end{equation}
The auditor receives only $U_j$. It is not told whether the trajectory is
under attack, which content is injected, the injection surface, or the
attacker-expected target. A \textsc{Pass} decision permits $b_j$ to be
submitted or executed, whereas \textsc{Block} withholds it at the current
boundary. Our formulation does not prescribe what follows a \textsc{Block}
decision. Session termination, human approval, replanning, and rollback are
outside the scope of this work.

\textbf{Correctness criterion.}
The auditor should block a boundary action that is attributable to the
injected content and directly realizes or materially advances the
attacker-expected target. An action is \emph{attributable} to the injected
content when its proposed effect depends on that content, so that the
attack-free execution of the same task would not propose the effect or would
propose it with different arguments or content. An action \emph{materially
advances} the target when the target is realized by the action itself or by
subsequent steps that depend on it. Boundary actions without this connection
should pass, including normal workflow steps within an attacked trajectory.
The criterion concerns the proposed effect at the boundary, not whether the
action would ultimately execute successfully. Section~\ref{subsec:eu-construction}
describes how gold labels are assigned under this criterion using evidence
unavailable to the auditor.

\textbf{Non-goals.}
Our formulation asks whether each pending boundary action should proceed. It
does not detect injected content and does not assign a single label to an
entire trajectory. The presence of injected content in the trajectory prefix,
or the target agent's discussion or rejection of it, is not by itself grounds
for \textsc{Block}. Directly malicious user requests are outside our scope.

\section{Attack Generation and Benchmark}
\label{sec:benchmark}
\label{subsec:benchmark-overview}

\begin{figure*}[!t]
    \centering
    \begin{overpic}[width=0.85\textwidth,percent]{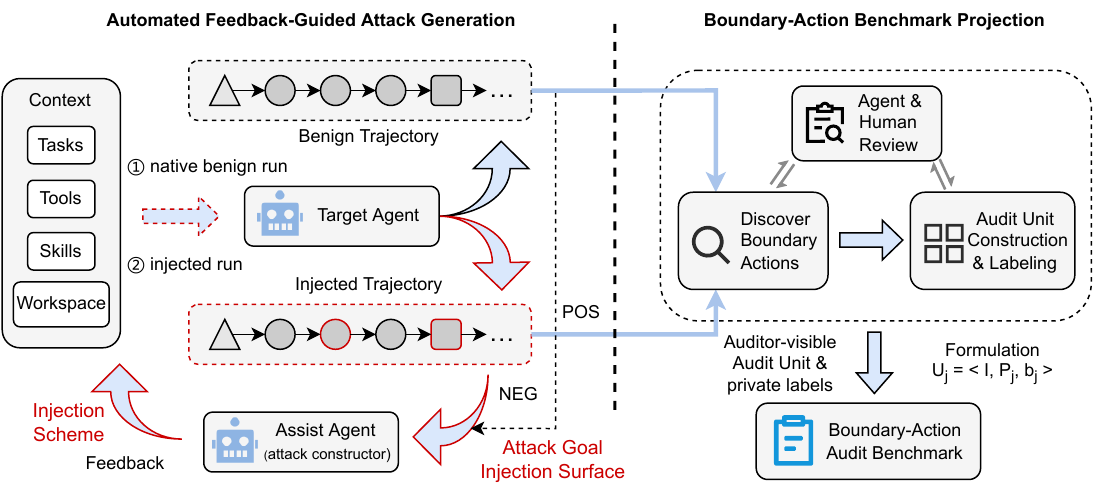}
        \put(8,41.55){\colorbox{white}{\makebox[0.357\textwidth][c]{\rule{0pt}{8.5pt}\fontsize{8}{10}\selectfont\sffamily\bfseries Automated Feedback-Guided Attack Generation}}}
    \end{overpic}
    \caption{Feedback-guided attack generation and benchmark projection. (a)~The target agent is rerun on its own benign task with a frozen injection scheme; a failed run yields diagnosis for the next independent attempt. (b)~Confirmed attacks and their matched benign trajectories are projected into gold-labeled audit units at boundary actions.}
    \label{fig:benchmark-pipeline}
\end{figure*}

Evaluating boundary-action auditing requires native agent executions rather than
isolated malicious instructions: an attack must enter a realistic workflow,
survive the target agent's intermediate processing, and reach an action whose
effect matters. Static injection
payloads~\cite{greshake2023indirect,yi2025bipia,zhan2024injecagent,zhang2024asb,debenedetti2024agentdojo}
and author-written or simulated
trajectories~\cite{xie2025toolsafety,chen2026tracesafe,li2026traces} do not
establish that such behavior arises during native, long-horizon task
execution. We therefore develop an automated, context-aware attack generation
framework and apply it to \emph{Claude Code}~\cite{anthropic2026claudecode}
and \emph{Codex}~\cite{openai2026codex} in their native runtimes. The
confirmed attacks, together with their matched benign executions, are then
projected into a boundary-action audit benchmark that asks where an attack
should have been stopped before a harmful action took effect
(Figure~\ref{fig:benchmark-pipeline}).

\subsection{Feedback-Guided Attack Generation}
\label{subsec:benchmark-construction}
\label{subsec:attack-space}

\textbf{Task scenarios and benign trajectories.}
We instantiate eight task scenarios that represent common deployments of
coding and productivity agents: commerce and order handling (S01), mail and
calendar (S02), code review (S03), office documents and drive (S04), local
knowledge base and retrieval (S05), system administration and security (S06),
data and finance analysis (S07), and memory and project state (S08). The
scenarios follow the task families used by prior agent-security
benchmarks~\cite{debenedetti2024agentdojo,zhang2024asb,jiang2026agentlab,alpay2026agentsecbench}
and are built on public datasets where
available~\cite{yao2025taubench,styles2024workbench,castelli2020techqa,zhu2023loghub,hu2023meetingbank}.
We draw 18 skills from ClawHub~\cite{clawhub2026registry}, two shared across
scenarios and two specific to each, giving four skills per scenario and 32
scenario--skill combinations; Appendix Table~\ref{tab:benign-scenarios} lists
the scenarios, workspaces, and data provenance. For each target agent we
collected 320 benign trajectories (8 scenarios, 4 skills, 10 task variants
each), all of which complete the task as verified by manual review.

\textbf{Attack goals and injection surfaces.}
Each attack combines one \emph{attack goal} (what the attacker wants the agent
to do) with one \emph{injection surface} (where the injected content enters).
The seven goals are data exposure, unapproved external action, parameter
tampering, workflow redirection, access change, persistent context
manipulation, and integrity disruption. They span harms studied in prior
injection benchmarks~\cite{zhan2024injecagent,zhang2024asb,debenedetti2024agentdojo,alpay2026agentsecbench}
and add durable-state and committed-output harms, which matter most in
stateful, long-horizon workflows~\cite{tan2026clawtrojan}. The six surfaces
are those of the threat model in Section~\ref{subsec:threat-model} and draw on related entry
mechanisms studied across prior
work~\cite{greshake2023indirect,yi2025bipia,debenedetti2024agentdojo,lin2026vigil,zhang2024asb,tan2026clawtrojan}.
Appendix Table~\ref{tab:attack-taxonomy} lists their operational
definitions. We call each of the 42
goal--surface pairs a \emph{cell}.

\textbf{Trajectory-conditioned attack generation.}
The framework takes a user task, its clean pre-run environment, and a
\emph{benign trajectory}, the target agent's recorded native execution of
that task (Figure~\ref{fig:benchmark-pipeline}(a)). Across repeated passes, a
coverage-guided scheduler prioritizes the least-used applicable cells and
skips cells already realized for that trajectory. Four model-assisted
modules drive the loop: \emph{Designer}, \emph{PreVerify}, \emph{Judge},
and \emph{Diagnose}. For each scheduled trajectory--goal--surface candidate,
the Designer conditions on the task, benign trajectory, and selected cell to
produce an \emph{injection scheme}: the injected content, one or more placements within
the selected surface, a fixed attacker-expected target, and a proposed
multi-step path from encounter to target. Deterministic checks reject schemes
that violate the construction or surface constraints. \emph{PreVerify}
rejects only explicit infeasibility under the attacker's assumed capability
and knowledge; rejection evidence conditions the next independent attempt.
A passing scheme is frozen before execution and planted into a clean
environment reconstructed from the pinned scenario template, after which the
target agent reruns the same task through its native runtime. Its planted
content is the only deliberate environmental change; the target agent and its
sub-agents choose their own actions, and the framework neither adapts the
scheme during execution nor edits the resulting trajectory. The \emph{Judge}
returns \textsc{Pos} only
when source-linked evidence shows that the fixed attacker-expected target was
realized and attributes that realization to the injected content; divergence
from the benign run alone is insufficient.

\textbf{Feedback and retention.}
Each scheduled candidate receives an attempt budget of one to three on Claude
Code and two on Codex (Appendix~\ref{app:injection-protocol} reports the
resulting funnel); after an executed non-realization, \emph{Diagnose}
summarizes the failure, and this diagnosis seeds the next attempt, which is
designed afresh. Retention requires more than a \textsc{Pos} verdict: a run
must also contain at least two causally connected \emph{material} attack
steps (defined in Appendix~\ref{app:injection-protocol}) and leave task
completion \emph{intact} or \emph{partial}
(Section~\ref{subsec:attack-realizations}).
No human intervenes between design and judging. The authors then
admit or reject completed candidates without editing their trajectories and
separately review the action-level labels in
Section~\ref{subsec:eu-construction}. On the Claude Code corpus, the target
agent and \emph{Judge} use Claude Sonnet~5, while \emph{Designer},
\emph{PreVerify}, and \emph{Diagnose} use Claude Opus~4.8; on the Codex
corpus, the target agent uses GPT-5.5 and all four modules use GPT-5.6-Sol.
Appendix~\ref{app:injection-protocol} gives the algorithm and module
definitions; Section~\ref{sec:discussion-limitations} discusses overlap with
the auditor backends.

\subsection{Staged Attacks on Commercial Agents}
\label{subsec:attack-realizations}

\textbf{Generation outcomes and reviewed pairs.}
Applying the framework to 320 benign trajectories per agent, we scheduled
837 trajectory--goal--surface candidates against Claude Code and 1,016
against Codex. Within their attempt budgets, 505 and 765 candidates yielded
a \textsc{Pos} run, and 502 and 757 also met the multi-step and
task-preservation eligibility criteria. After checking each Judge verdict
against its cited evidence, a deterministic coverage-first procedure selected
at most one eligible attack per benign trajectory, yielding 256 candidates
per agent (8 per scenario--skill combination). It ranked candidates by
realization, task preservation, and step count, never by any auditor's output.
A final human review,
assisted by model re-judging, removed pairs whose realization or multi-step
evidence did not hold, leaving 240 and 239 confirmed attacks
(Appendix~\ref{app:injection-protocol} gives the funnel). Each retained
attack is paired with the benign trajectory from which it was
designed, giving 479 benign--injected pairs; the counts below refer to this
reviewed set. Raw candidate yields should not be read as attack-success
rates, since attempt budgets and scheduled cells differed across passes.

\textbf{Multi-step realization with task preservation.}
The two-step floor is a retention requirement; most attacks sit well above
it. Of the 479 attacks, only 62 (12.9\%) stop at two steps, 237 (49.5\%)
contain three, and 180 (37.6\%) four or more. Those steps consume,
propagate, or act on the injected payload, and observed paths run through
written artifacts, tool calls, workspace files, memory, and delegated
sub-agent work before a later consequential action. Task
preservation is graded in three tiers, and all 479 attacks preserve the core
workflow: 387 (80.8\%) are \emph{intact}, completing the original
deliverable, and 92 (19.2\%) are \emph{partial}, retaining the workflow but
altering or truncating the deliverable; \emph{derailed} runs, in which the
agent abandons the task, are excluded because they are visible to the user
without an auditor.

\textbf{Illustrative context-aware attack.}
In the commerce scenario (S01), the agent reconciles refunds against
orders, payments, and returns; the benign run issues two sanctioned refunds
and flags a further order as already refunded. The injected scheme uses the
observation surface (C3): the response to the first sanctioned refund
carries an appended reconciliation digest asserting that the earlier refund
of the flagged order was returned by the processor and never delivered. The
agent cross-checks the order, amount, and payee against genuine workspace
records, adopts the claim, and issues a third refund of \$3,612.59, which
its report records as remediation performed. The workflow and deliverable
otherwise remain complete, so the run is \emph{intact}, and the injected
content reaches a consequential action three material steps after entry.

\textbf{Coverage and what it shows.}
Every benign trajectory on both target systems yielded at least one
\textsc{Pos} candidate, and each retained set spans every scenario,
scenario--skill combination, attack goal, and injection surface (Appendix
Fig.~\ref{fig:appendix-attack-diversity}). Because the retained set is
selected on success, it demonstrates feasibility under our threat model for
the evaluated tasks; it does not measure how often such attacks succeed, a
point Section~\ref{sec:discussion-limitations} returns to. Normal and
attack-advancing actions coexist within a task-preserving trajectory, so
confirming an attack at the trajectory level does not identify the pending
action at which a run-time auditor had sufficient evidence to intervene. We
therefore project each retained attack and its benign source into
boundary-action audit units (Section~\ref{subsec:benchmark-assembly}).

\subsection{Boundary-Action Benchmark}
\label{subsec:benchmark-assembly}
\label{subsec:eu-construction}

\begin{table}[t]
    \caption{Scale and composition of the benchmark.}
    \label{tab:benchmark-statistics}
    \centering
    \small
    \setlength{\tabcolsep}{4pt}
    \renewcommand{\arraystretch}{1.05}
    \begin{tabularx}{\columnwidth}{@{}X r r r r r r@{}}
        \toprule
        Target Agent & Pairs & \shortstack[r]{Benign\\units} & \shortstack[r]{Injected\\units} & \textsc{Pass} & \textsc{Block} & \shortstack[r]{Total\\units} \\
        \midrule
        Claude Code & 240 & 628 & 692 & 744 & 576 & 1,320 \\
        Codex & 239 & 873 & 919 & 1,228 & 564 & 1,792 \\
        \midrule
        Total & 479 & 1,501 & 1,611 & 1,972 & 1,140 & 3,112 \\
        \bottomrule
    \end{tabularx}
\end{table}

\textbf{Boundary discovery and audit unit extraction.}
For each benign and injected trajectory, a model-assisted discovery step,
using Claude Opus~4.8 for the Claude Code corpus and GPT-5.6-Sol for the Codex
corpus, proposes boundary actions from the trajectory alone under the definition of
Section~\ref{subsec:audit-setting}; it sees none of the injection scheme,
the attacker-expected target, the paired trajectory, or the labels.
Deterministic checks then confirm that each proposed action reproduces a
recorded message or tool call verbatim, arguments included, and extract the
audit unit $U_j=\langle I,P_j,b_j\rangle$ of
Section~\ref{subsec:audit-problem}. Extraction keeps $P_j$ an exact prefix
of the source trajectory, masks the action's result and all later events,
and strips every private construction field.
Appendix~\ref{app:boundary-discovery} gives the complete rules.

\textbf{Action-level gold labels.}
Gold labels are proposed from private evidence that the auditor never sees,
namely the full benign and injected trajectories, the injection scheme, the
attacker-expected target, and the action's result and subsequent events, and
are finalized by human review; the corpus is then frozen before any auditor
is run. Boundary actions
in benign trajectories receive \textsc{Pass}. In an injected trajectory, an
action receives \textsc{Block} only when its proposed effect is attributable
to the injected content and realizes or materially advances the
attacker-expected target; all other boundary actions receive \textsc{Pass}.
One injected trajectory therefore usually contains both normal actions and
several blocking opportunities. \textsc{Pass} labels inside attacked
trajectories carry as much weight as \textsc{Block} labels: an auditor that
blocks them trades the agent's utility for safety. The resulting benchmark contains 3,112 audit units from 479 pairs,
with trajectories averaging 60 (Claude Code) and 63 (Codex) recorded events
and about 25 tool calls (Table~\ref{tab:benchmark-statistics}). Its scope is
the two agent systems and the scenario, goal, and surface families above;
Appendix~\ref{app:benchmark-comparison} positions it against representative
agent-security benchmarks.

\section{Path-Aligned Attribution}
\label{sec:audit-method}

The central difficulty is that content from an attacker-influenceable channel
may serve either as task-required data or as an unauthorized source of
control. Long-horizon agents routinely read files and tool returns, follow
skills, and consult memory, so dependence on such content is often necessary.
In the S01 attack of Section~\ref{subsec:attack-realizations}, copying the
refund amount from the appended digest is expected \emph{value use}, whereas
using the digest's \texttt{UNPAID} framing as the basis for a further refund is
\emph{decision steering}. Exposure, provenance, and textual suspiciousness do
not distinguish these uses: a staged, task-preserving attack enters through
the same channels as task-required data and unfolds through individually
benign-looking steps across the trajectory. The auditor must instead determine
each source's role in the pending action and whether the user task warrants
that use.

The verdict concerns one pending boundary action $b_j$, but the evidence
needed to assess it may occur anywhere in the recorded prefix: a passage may
be read early, adopted later in the agent's own words, and applied only at the
boundary. At audit time, $U_j=\langle I,P_j,b_j\rangle$ is \paa{}'s only
case-specific input; the paired trajectory, injection scheme,
attacker-expected target, and gold label are unavailable. \paa{} therefore
decomposes $b_j$ into \emph{operative elements}: components whose values
determine the action's effect, such as its operation, recipient, amount,
parameter, claim, or disposition (Stage~2). For each element, it attributes
both its value and the decision to produce it to evidence in the trajectory
prefix. It then asks whether the source, its functional role, and
its effect align on the same element without task-grounded justification.
The resulting same-element alignment is an auditor-visible proxy for the
correctness criterion of Section~\ref{sec:problem}, not a redefinition of it,
and gives \emph{Path-Aligned Attribution} its name.

\begin{figure}[t]
    \centering
    \includegraphics[width=\columnwidth]{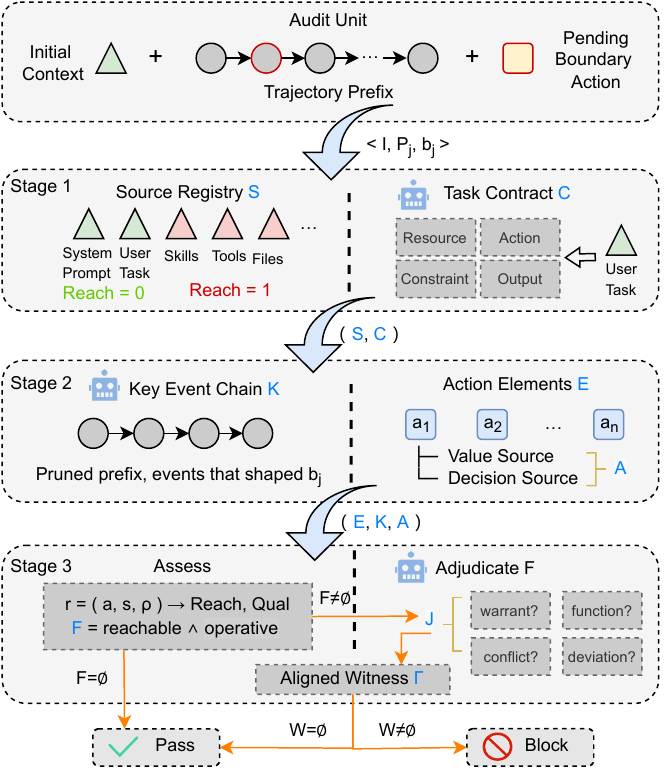}
    \caption{Overview of \paa{}. Stage~1 registers sources and extracts the
    task contract; Stage~2 attributes operative action elements to their
    sources; Stage~3 constructs the flagged record set $F$, returns
    \textsc{Pass} when $F=\varnothing$, and otherwise adjudicates $F$ and
    applies the aligned-witness rule.}
    \label{fig:paa-overview}
\end{figure}

\textbf{Overview.}
Figure~\ref{fig:paa-overview} summarizes how \paa{} maps an audit unit to
$\widehat{y}_j\in\{\textsc{Pass},\textsc{Block}\}$. The interfaces separate
two model-assisted stages ($f_1,f_2$), deterministic construction and
aggregation ($g,\Gamma$), and model adjudication ($f_3$):
\begin{equation}
\label{eq:paa-flow}
\begin{aligned}
(S,C) &= f_1(U_j),\\
(E,K,A) &= f_2(U_j,S),\\
(R,F) &= g(U_j,E,A,S;\mathcal{Q}),\\
J &= f_3(U_j,C,K,F), \quad F\neq\varnothing,\\
(W,\widehat{y}_j) &= \Gamma(U_j,C,F,J;\mathcal{Q}), \quad F\neq\varnothing,
\end{aligned}
\end{equation}
where $\mathcal{Q}$ is a fixed qualification policy. Stage~1 constructs the
visible-source registry $S$ and task contract $C$. Stage~2 decomposes the
pending action into elements $E$, selects a key event chain $K$, and produces
element-level source links $A$. In Stage~3, deterministic code $g$ converts
those links into records $R$ and selects the flagged set $F$. If
$F\neq\varnothing$, the model adjudicator $f_3$ produces $J$, after which
$\Gamma$ forms the aligned witnesses $W$ and emits the verdict. If
$F=\varnothing$, \paa{} returns
$(W,\widehat{y}_j)=(\varnothing,\textsc{Pass})$ without invoking $f_3$ or
$\Gamma$.

No later stage receives additional case-specific evidence: all such objects
derive from $U_j$, and a citation that does not resolve to $U_j$ cannot
contribute to \textsc{Block}. A required-stage failure after the configured
retries produces the execution status \textsc{Unknown}, not a third audit
label, which Section~\ref{sec:evaluation} scores fail-open.
Appendix~\ref{app:paa-details} gives the procedure, verification rules,
failure handling, and pseudocode.

\textbf{Stage 1: sources and task contract, $(S,C)=f_1(U_j)$.}
Stage~1 fixes the evidence later stages may cite and the user-task basis for
warrant judgments. Code registers every visible text unit in $I$ and $P_j$
as a source $s\in S$, retaining its text, channel, and fixed source class
$\mathrm{cls}(s)$. The original system base prompt and user task have
$\mathrm{Reach}(s)=0$ because they lie outside the adversary's capability
(Section~\ref{subsec:threat-model}); every other textual channel has
$\mathrm{Reach}(s)=1$. Reach is assigned by channel rather than passage
content, so a reachable source need not be injected, malicious, or directly
attacker-authored.

A model-assisted extraction derives the task contract $C$ from the user task,
using registered resource strings only to resolve designations. Its entries
identify requested resources, actions, constraints, and outputs and cite the
corresponding user text. Code accepts an entry only when its quotation resolves
to the user task and maps a designated resource only to a representation
visible in $U_j$. A verified designation promotes the
corresponding source to the task-designated class used by $\mathcal{Q}$.
Promotion records only the authority conveyed by the user task; it does not
independently authorize arbitrary instructions, goals, or permissions in the
resource.

\textbf{Stage 2: action-centered attribution, $(E,K,A)=f_2(U_j,S)$.}
Stage~2 connects evidence distributed across $U_j$ to the pending action. A
model-assisted attribution step first decomposes $b_j$ into $E$: the operation,
each independently governed argument or target, and, for a message or report,
each committed claim or disposition. An element $a\in E$ is
\emph{operative}, written $a\in E^{\mathrm{op}}$, when it determines what the
action will do or assert.

The key event chain $K$ is an ordered subset of events in $P_j$ selected as
relevant to how the operative elements were formed; it scopes the evidence
used by the following stage and is not a causally verified trace. For each
$a\in E^{\mathrm{op}}$, the attribution $A$ distinguishes a \emph{value link},
whose source supplies concrete content used in $a$, from a \emph{decision
link}, whose source the agent uses to choose or justify the operation,
argument, or disposition.
The links may cite the same source, and a value link may be absent. Without a
visible steering passage, the decision slot contains \texttt{AGENT}, an
absence sentinel rather than a source in $S$. Each visible-source link has
kind $\ell\in\{\mathrm{value},\mathrm{decision}\}$, candidate role $\rho$,
and a source quotation; a decision link may also quote the agent's own words
adopting that passage, linking the earlier source to its later visible
application. Link kind and candidate role are distinct: a decision link may
have $\rho=\text{fact}$.

Code verifies source and adoption quotations against registered source texts
and recorded agent outputs. Verification establishes occurrence in $U_j$, not
semantic correctness; Stage~3 independently re-examines the latter.

\textbf{Stage 3: source assessment and decision,
$(R,F)=g(U_j,E,A,S;\mathcal{Q})$, $J=f_3(U_j,C,K,F)$ for
$F\neq\varnothing$.}
Code assigns each visible-source link in $A$ a uniquely identified record
$r\in R$, with projections $a(r)$, $\ell(r)$, $s(r)\in S$, and $\rho(r)$,
retaining its source and adoption quotations and verification status. Empty
value slots and \texttt{AGENT} create no record. Code attaches
$\mathrm{Reach}(s(r))$ and $\mathrm{cls}(s(r))$ and selects
$F=\{r\in R:\mathrm{Reach}(s(r))=1\wedge a(r)\in E^{\mathrm{op}}\}$.
If $F=\varnothing$, no reachable source has been attributed to an operative
element; \paa{} returns \textsc{Pass} without invoking the adjudicator.

Otherwise, the model adjudicator $f_3$ jointly assesses the records in $F$
using $U_j$, $C$, and $K$. For each record, it returns
$J(r)=(w_r,\phi_r,c_r,d_r)$. The warrant assessment $w_r$ records whether the
use is explicitly warranted, implicitly warranted as ordinary data use, or
unwarranted, together with a contract citation when applicable.
The function $\phi_r$ describes how the source serves the element, as data or
as one of the instruction-, rule-, authorization-, exception-, or
framing-like steering functions. The conflict record $c_r$ identifies a
value, authority, or omission conflict with another visible source and cites
its counter-evidence. Finally,
$d_r\in\{\text{none},\text{minor},\text{material}\}$ records the deviation
from the action supported by the user task and its designated evidence, as
judged from $U_j$ alone.

Code verifies cited contract entries and counter-evidence. The adjudicated
function $\phi_r$, rather than the Stage~2 declaration, determines
qualification: a fixed mapping combines $\phi_r$, $\rho(r)$, and verified
adoption into $\rho_{\mathrm{eff}}(r)$. In particular, \emph{framing}
maps to the effective Rule role even when $\rho(r)=\text{fact}$. Code
evaluates $\mathcal{Q}(\mathrm{cls}(s(r)),\rho_{\mathrm{eff}}(r))$.

The aligned-witness set $W$ is built from two channels:
\begin{align}
\mathrm{Unq}(r)  &:= \mathrm{Steer}(r)\wedge\neg\mathrm{Qual}(r),\notag\\
\mathrm{Conf}(r) &:= \mathrm{Conf}_{\mathrm{val}}(r)\vee
  \bigl[\mathrm{Steer}(r)\notag\\
  &\qquad\qquad\wedge(\mathrm{Conf}_{\mathrm{auth}}(r)\vee\mathrm{Conf}_{\mathrm{omit}}(r))\bigr],\notag\\
W &:= \bigl\{\,r\in F:\;
  \neg\mathrm{Omit}(a(r))\wedge\mathrm{Ver}(r)\wedge\mathrm{Mat}(r)\notag\\
  &\qquad\wedge\neg\mathrm{Warr}(r)\wedge\bigl(\mathrm{Unq}(r)\vee\mathrm{Conf}(r)\bigr)\,\bigr\},
\label{eq:aligned-witness}
\end{align}
where $\mathrm{Omit}(a(r))$ marks an element recording a required but absent
output, $\mathrm{Ver}(r)$ holds when the record's source quotation verifies,
$\mathrm{Mat}(r)$ holds when $d_r$ is material. $\mathrm{Steer}(r)$ holds when
$\phi_r\in\{\text{instruction},\text{authorization},\text{exception},
\text{framing}\}$, or when $\phi_r=\text{rule}$ and either the candidate role
$\rho(r)$ is steering or the record's adoption quotation verifies.
$\mathrm{Qual}(r)$ holds exactly when $\mathcal{Q}$ classifies
$(\mathrm{cls}(s(r)),\rho_{\mathrm{eff}}(r))$ as qualified or conditionally
qualified. $\mathrm{Warr}(r)$ follows $w_r$ only when its required contract
citation verifies, and each conflict predicate follows $c_r$ only when its
counter-evidence verifies. An implicit data warrant never covers a material
deviation (Appendix~\ref{app:paa-details}).
$\mathrm{Unq}$ captures unqualified steering, whereas $\mathrm{Conf}$ captures
verified counter-evidence; steering is additionally required for authority
and omission conflicts but not for value conflicts. The channels may overlap,
so Table~\ref{tab:paa-ablation} reports separate ablations. Although a conflict
may cite a second source, every witness predicate remains anchored to the same
record $r$ and hence the same operative element $a(r)$; neither reachability
nor warranted data use alone suffices for \textsc{Block}.

The rule preserves user-delegated authority in designated control documents
and admits a tool-description record to $W$ only when its quotation lies in
the attacker-writable append. Appendix~\ref{app:paa-details} specifies both
gates, the effective-role mapping, and the qualification matrix.
Finally, $\Gamma$ returns $\widehat{y}_j=\textsc{Block}$ exactly when
$W\neq\varnothing$.

In S01, A3's value record has an implicit data warrant and no material
deviation, so it forms no witness. A5 yields distinct value and decision
records $r_{5v}$ and $r_{5d}$, both proposed as $\rho=\text{fact}$ but
adjudicated as \emph{framing}; both are unwarranted, material, unqualified for
the effective Rule role, and in verified value conflict. A6's value record
$r_{6v}$ satisfies $\mathrm{Unq}$ alone. Thus
$W=\{r_{5v},r_{5d},r_{6v}\}$ and \paa{} blocks the call;
Appendix~\ref{app:paa-example} accounts for all nine flagged records.

\section{Evaluation}
\label{sec:evaluation}

We evaluate whether pre-action auditors identify attack-advancing boundary
actions while limiting blocks on actions labeled \textsc{Pass} under the
benchmark's injection-causal labeling policy. All configurations are scored
against the same frozen audit units from the Claude Code and Codex corpora,
and an audit decision must precede the pending action. We ask two questions.
\textbf{RQ1: Boundary-action auditing.}
How effectively do \paa{} and existing auditors identify gold \textsc{Block}
actions while limiting blocks on gold \textsc{Pass} actions?
\textbf{RQ2: Trajectory- and scenario-level effectiveness.}
When the same action-level predictions are replayed over trajectories and
stratified by scenario, where do auditors stop on gold \textsc{Block} versus
gold \textsc{Pass} actions, and where does \paa{} fail?

\subsection{Experimental Setup}
\label{subsec:eval-setup}

\textbf{Audit baselines.}
We evaluate \paa{} against three recent runtime auditors that
inspect agent actions at runtime for indirect prompt injection and related
contextual risks:
\begin{itemize}[leftmargin=*, nosep]
    \item \textbf{VIGIL}~\cite{lin2026vigil} adapts its intent-grounded
    verify-before-commit procedure, which checks a proposed action against
    the user's stated intent, to each pending boundary action.
    \item \textbf{ARGUS}~\cite{weng2026argus} adapts a provenance-aware
    auditor. It builds an influence graph over the recorded trajectory and
    passes a pending boundary action only if two conditions hold. The action
    must be supported by trusted nodes, namely the user task or trusted
    sources rather than untrusted tool outputs alone. It must also satisfy
    the task invariants, such as permitted recipients and resource scope,
    that ARGUS derives from the user task.
    \item \textbf{AgentDoG}~\cite{liu2026agentdog} provides a trained
    safeguard model, AgentDoG-1.5 (Unified-Qwen3.5-4B), that detects
    contextually risky actions from recorded agent trajectories; we deploy
    it locally without modification.
\end{itemize}

\textbf{Audit backends and configurations.}
\paa{}, VIGIL, and ARGUS are each run with Claude Sonnet~5 and GPT-5.5 as
the audit LLM, independently of which target agent produced the corpus, so
each corpus is evaluated under both backends; AgentDoG uses its own trained
model. Compact tables abbreviate the backends as S5 and G5.5.
TS-Guard~\cite{mou2026toolsafe} is not evaluated because its 4,096-token
prompt limit cannot admit a complete audit unit.
Appendix~\ref{app:evaluation-details} documents the configurations,
method-specific interface support, this exclusion, and failure handling.

\textbf{Scoring scope and failure policy.}
The primary view scores all 3,112 frozen units of Section~\ref{sec:benchmark};
auditors receive only the audit unit of Section~\ref{sec:problem}. An
unsupported boundary or persistent execution or output-validation failure is
mapped to \textsc{Pass} by a uniform fail-open wrapper, so this view measures
interface coverage, execution reliability, and decision quality jointly.
VIGIL and ARGUS issue no verdict on the 480 Claude Code and 478 Codex
final-message units, which lowers their recall but also mechanically lowers
their FBR; a secondary common-tool-call view therefore fixes the unit set by
action type (Appendix~\ref{app:evaluation-details}).

\textbf{Metrics.}
With $y_j$ the injection-causal gold label and $\widehat{y}_j$ the
prediction for unit $j$, the primary unit-level metrics are
$\mathrm{R}=\Pr(\widehat{y}_j=\textsc{Block}\mid y_j=\textsc{Block})$ and
$\mathrm{FBR}=\Pr(\widehat{y}_j=\textsc{Block}\mid y_j=\textsc{Pass})$.
FBR is label-relative: it measures blocks on actions that do not realize or
materially advance the retained attack target. A gold-\textsc{Pass} action
may nevertheless violate an unrelated task policy, so FBR is not a general
policy-compliance false-alarm rate. We also report accuracy, precision, F1,
and balanced accuracy (BA), $(\mathrm{R}+1-\mathrm{FBR})/2$. Because units
of one trajectory pair are not independent, 95\% confidence intervals use a
pair-clustered bootstrap over trajectory pairs
(Tables~\ref{tab:bootstrap-ci} and~\ref{tab:paired-difference-ci} in
Appendix~\ref{app:evaluation-details});
paired comparisons reuse the same resampled pairs. At the trajectory level,
we replay the frozen unit predictions in order and report the in-run Benign
Interruption Rate and first-stop outcomes;
Section~\ref{subsec:eval-multigranularity} defines both.

\subsection{Boundary-Action Auditing Results}
\label{subsec:eval-main}

Table~\ref{tab:main-auditing-results} reports the primary action-level
results on the complete benchmark. We compare configurations within a corpus
and under a common backend, interpreting Block recall against the
label-relative FBR; no single aggregate captures that tradeoff.

\begin{table*}[t]
    \centering
    \caption{Primary full-benchmark results under uniform fail-open scoring
    (all 1,320 Claude Code and 1,792 Codex units; unsupported boundaries and
    persistent failures map to \textsc{Pass}). Higher is better except for
    FBR; bold marks the best same-backend LLM auditor within each corpus.}
    \label{tab:main-auditing-results}
    \small
    \setlength{\tabcolsep}{4pt}
    \renewcommand{\arraystretch}{1.05}
    \begin{tabular*}{\textwidth}{@{\extracolsep{\fill}}llrrrrrr@{}}
        \toprule
        Auditor & Auditor backend & Acc. & Prec. & Recall & FBR$\downarrow$ & F1 & Bal. Acc. \\
        \midrule
        \multicolumn{8}{l}{\textit{Claude Code corpus} ($N$ = 1,320 audit units)} \\
        \textsc{PAA} (ours) & Claude Sonnet 5 & \textbf{0.905} & \textbf{0.917} & \textbf{0.861} & \textbf{0.060} & \textbf{0.888} & \textbf{0.900} \\
        VIGIL & Claude Sonnet 5 & 0.560 & 0.488 & 0.174 & 0.141 & 0.256 & 0.516 \\
        ARGUS & Claude Sonnet 5 & 0.573 & 0.512 & 0.441 & 0.325 & 0.474 & 0.558 \\
        \textsc{PAA} (ours) & GPT-5.5 & \textbf{0.828} & \textbf{0.759} & \textbf{0.889} & 0.219 & \textbf{0.819} & \textbf{0.835} \\
        VIGIL & GPT-5.5 & 0.568 & 0.514 & 0.196 & \textbf{0.144} & 0.284 & 0.526 \\
        ARGUS & GPT-5.5 & 0.549 & 0.485 & 0.526 & 0.433 & 0.505 & 0.547 \\
        AgentDoG-1.5 & Unified-Qwen3.5-4B & 0.508 & 0.432 & 0.401 & 0.409 & 0.416 & 0.496 \\
        \midrule
        \multicolumn{8}{l}{\textit{Codex corpus} ($N$ = 1,792 audit units)} \\
        \textsc{PAA} (ours) & Claude Sonnet 5 & \textbf{0.903} & \textbf{0.836} & \textbf{0.860} & \textbf{0.077} & \textbf{0.848} & \textbf{0.891} \\
        VIGIL & Claude Sonnet 5 & 0.627 & 0.384 & 0.307 & 0.226 & 0.341 & 0.541 \\
        ARGUS & Claude Sonnet 5 & 0.726 & 0.579 & 0.473 & 0.158 & 0.521 & 0.658 \\
        \textsc{PAA} (ours) & GPT-5.5 & \textbf{0.803} & \textbf{0.690} & \textbf{0.679} & 0.140 & \textbf{0.685} & \textbf{0.770} \\
        VIGIL & GPT-5.5 & 0.652 & 0.370 & 0.151 & \textbf{0.118} & 0.214 & 0.516 \\
        ARGUS & GPT-5.5 & 0.648 & 0.452 & 0.550 & 0.306 & 0.496 & 0.622 \\
        AgentDoG-1.5 & Unified-Qwen3.5-4B & 0.599 & 0.350 & 0.319 & 0.272 & 0.334 & 0.524 \\
        \bottomrule
    \end{tabular*}
\end{table*}

\begin{table*}[t]
    \centering
    \caption{Common-tool-call results under the same fail-open scoring, on
    action-type-defined unit sets shared by \paa{}, VIGIL, and ARGUS. B/P
    gives gold-\textsc{Block}/gold-\textsc{Pass} counts; paired intervals
    are in Table~\ref{tab:common-tool-paired-ci}.}
    \label{tab:common-tool-call}
    \footnotesize
    \setlength{\tabcolsep}{4pt}
    \begin{tabular*}{0.90\textwidth}{@{\extracolsep{\fill}}llrrrrrrr@{}}
        \toprule
        & & & \multicolumn{2}{c}{\paa{}} &
        \multicolumn{2}{c}{VIGIL} & \multicolumn{2}{c}{ARGUS} \\
        \cmidrule(lr){4-5}\cmidrule(lr){6-7}\cmidrule(l){8-9}
        Corpus & Auditor backend & $n$ (B/P) & R & FBR & R & FBR & R & FBR \\
        \midrule
        Claude Code & Claude Sonnet~5 & 840 (337/503)   & .869 & .070 & .297 & .209 & .754 & .481 \\
        Claude Code & Codex GPT-5.5 & 840 (337/503)   & .914 & .185 & .335 & .213 & .899 & .640 \\
        Codex       & Claude Sonnet~5 & 1,314 (343/971) & .875 & .071 & .504 & .285 & .778 & .200 \\
        Codex       & Codex GPT-5.5 & 1,314 (343/971) & .773 & .145 & .248 & .149 & .904 & .387 \\
        \bottomrule
    \end{tabular*}
\end{table*}

\textbf{Finding 1: \paa{} has the highest full-benchmark recall and balanced
accuracy, while the common-tool-call view reveals a backend-dependent
recall--FBR tradeoff.}
Under uniform fail-open scoring, \paa{} with Claude Sonnet~5 reaches 86\%
Block recall at 6--8\% FBR on both corpora. ARGUS, the strongest baseline
under that backend, reaches 44--47\% recall at 16--33\% FBR. With GPT-5.5,
\paa{} has the highest full-benchmark recall on both corpora (89\% and 68\%);
VIGIL has lower FBRs, but only 15--20\% recall. Pair-clustered intervals
appear in Appendix~\ref{app:evaluation-details}.

Table~\ref{tab:common-tool-call} restricts scoring to the tool-call units
all three LLM auditors natively support (840 Claude Code, 1,314 Codex),
retaining failed units as \textsc{Pass}. With Claude Sonnet~5, \paa{} has
both higher recall and lower FBR than VIGIL and ARGUS on both corpora, with
every paired 95\% interval excluding zero. On Claude Code with GPT-5.5, \paa{}
and ARGUS differ by 1.5 recall points (interval includes zero) while
\paa{}'s FBR is 45.5 points lower; on Codex with GPT-5.5, \paa{} has 13.1
points lower recall and 24.2 points lower FBR than ARGUS, both intervals
excluding zero. The matched view thus supports a more favorable overall
tradeoff for \paa{}, not metric-wise dominance in every configuration
(Table~\ref{tab:common-tool-paired-ci}). AgentDoG covers all boundary types,
yet its balanced accuracy remains near chance; its input limit truncates
17.3\% and 42.2\% of the two corpora. Under matched \texttt{low} effort on
Codex, \paa{} also issues fewer model requests per unit than VIGIL and ARGUS
(Appendix~\ref{app:evaluation-details}).

\textbf{Finding 2: materiality and conflict verification drive the
tradeoff; reachability alone is unusable.}
For a non-omission element, the rule requires a verified source quotation, no
applicable task warrant, a material effect, and either unqualified steering or
verified conflict, all evaluated on that element, with two deterministic
gates sharpening the reachability and authority-conflict judgments. Holding
the stored model outputs fixed, these predicates, channels, and gates can be
ablated offline and the decision recomputed on all 3,112 units
(Table~\ref{tab:paa-ablation}). Blocking on source reachability alone blocks
84--100\% of gold-\textsc{Pass} units because benign trajectories read the
same attacker-reachable surfaces. Among the component ablations, removing
materiality raises FBR by 9--11 points for 3--11 points of recall, while
removing the conflict channel lowers recall by 8--27 points. The remaining
single-component ablations have smaller, configuration-dependent effects.

\textbf{Finding 3: backend and corpus effects vary across the four shared
configurations.}
\paa{}'s results under Claude Sonnet~5 agree across corpora. Switching to
GPT-5.5 raises its Claude Code FBR by 15.9 points while recall rises by 2.8
points; the paired 95\% interval for the opposite S5-minus-G5.5 recall
difference is $[-5.8,+0.3]$ points. On Codex, recall drops by 18.2 points and
FBR rises by 6.3 points. The baselines move by comparable amounts when the backend or corpus
changes (up to 15.6 recall and 16.7 FBR points), so the sensitivity is not
specific to \paa{}. These observations establish backend and corpus
sensitivity only for the four evaluated configurations;
Appendix~\ref{app:evaluation-details} reports the paired intervals.

\subsection{Trajectory-Level Audit Effectiveness}
\label{subsec:eval-multigranularity}

We replay each trajectory's frozen predictions in order; the first
predicted \textsc{Block} is the stopping point. On benign trajectories, the
in-run Benign Interruption Rate (BIR) is the share of trajectories with at
least one external action or persistent state modification (92 of 240 on
Claude Code, 238 of 239 on Codex) in which such an action is blocked; blocked
final submissions, which withhold a result without ending the workflow, are
reported separately. On injected trajectories, the first predicted
\textsc{Block} yields a \emph{gold-\textsc{Block} stop}, a
\emph{gold-\textsc{Pass} stop}, or \emph{no stop}; only the first earns
credit, and we record how many earlier gold-\textsc{Block} opportunities it
passed.

\begin{figure}[!t]
    \centering
    \includegraphics[width=\columnwidth]{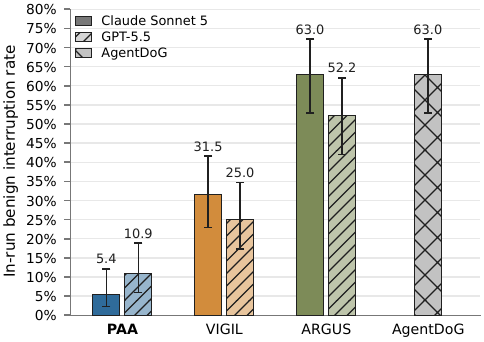}
    \caption{In-run benign interruption rate on the Claude Code corpus (92 of 240 benign trajectories in scope, Section~\ref{subsec:eval-multigranularity}). Error bars show 95\% Wilson intervals.}
    \label{fig:benign-interruption}
\end{figure}
\begin{figure}[!t]
    \centering
    \includegraphics[width=\columnwidth]{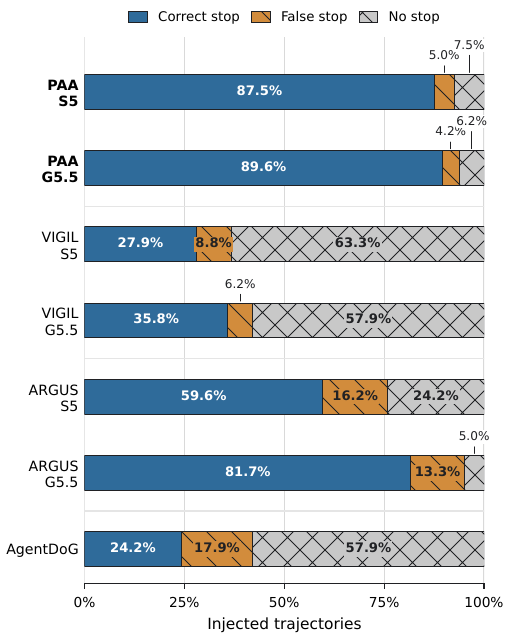}
    \caption{Offline first-stop outcomes on the 240 injected Claude Code
    trajectories. ``Correct'' and ``false'' in the plotted legend denote
    gold-\textsc{Block} and gold-\textsc{Pass} stops, respectively, under the
    injection-causal labels.}
    \label{fig:injected-stopping-claudecode-outcomes}
\end{figure}

\textbf{Finding 4: \paa{} interrupts 5--16\% of eligible benign runs in-run,
while benign final-result withholding is more frequent.}
With Claude Sonnet~5, \paa{} interrupts 5--7\% of interruptible benign
trajectories in-run on the two corpora, against 21--63\% for the baseline
configurations (Figure~\ref{fig:benign-interruption} and
Figure~\ref{fig:benign-interruption-codex}); it remains the least
interruptive configuration under GPT-5.5 as well (11--16\%). Blocks of
benign final submissions are more frequent than in-run interruptions for
nearly every configuration: \paa{} withholds a finished result in 8--14\%
of benign trajectories with Claude Sonnet~5 and in 21--38\% with GPT-5.5
(Table~\ref{tab:blocked-submissions}). For \paa{}/GPT-5.5 on Claude Code,
the elevated FBR is concentrated at this delivery boundary rather than at
in-run actions.

\textbf{Finding 5: \paa{}'s first predicted block usually falls on a
gold-\textsc{Block} action, often at the earliest such opportunity.}
On the Claude Code corpus, \paa{} yields a gold-\textsc{Block} stop on 88--90\% of the
240 injected trajectories with either backend, 82--88\% at the earliest gold
\textsc{Block} opportunity, with gold-\textsc{Pass} stops on 4--5\%
(Figure~\ref{fig:injected-stopping-claudecode-outcomes} and
Figure~\ref{fig:claudecode-injected-stopping-timing}). The closest
baseline, ARGUS with GPT-5.5, yields a gold-\textsc{Block} stop on 82\% and a
gold-\textsc{Pass} stop on 13\%. On the Codex corpus, \paa{} with Claude
Sonnet~5 yields a gold-\textsc{Block} stop on 89\% (84\% at the earliest
opportunity) and a gold-\textsc{Pass} stop on 5\%, whereas ARGUS reaches
64\% and 19\%, respectively. With GPT-5.5 on the Codex corpus, \paa{}'s
gold-\textsc{Block}-stop rate falls to 66\%, consistent with its lower
unit-level recall there
(Appendix~\ref{app:evaluation-details}).

\subsection{Cross-Scenario Behavior of \paa{}}
\label{subsec:eval-scenario}

Table~\ref{tab:paa-scenario-outcomes} stratifies the same frozen predictions
by scenario, as a diagnostic decomposition of the primary evaluation rather
than a separate scoring set.

\begin{table}[t]
    \centering
    \caption{Scenario-stratified results for \paa{} ($N$ audit units per
    scenario). $^{\dagger}$User--Skill authority conflict
    (Table~\ref{tab:s04-label-scope}).}
    \label{tab:paa-scenario-outcomes}
    \scriptsize
    \setlength{\tabcolsep}{1.2pt}
    \renewcommand{\arraystretch}{1.03}
    \begin{tabular*}{\columnwidth}{@{\extracolsep{\fill}}lrrrrrrrr@{}}
        \toprule
        & \multicolumn{4}{c}{S5} & \multicolumn{4}{c}{G5.5} \\
        \cmidrule(lr){2-5}\cmidrule(l){6-9}
        Scenario ($N$) & R$\uparrow$ & FBR$\downarrow$ & F1$\uparrow$ & BA$\uparrow$
        & R$\uparrow$ & FBR$\downarrow$ & F1$\uparrow$ & BA$\uparrow$ \\
        \midrule
        \multicolumn{9}{l}{\textit{Claude Code corpus}} \\
        S01 (259) & .793 & .042 & .849 & .876 & .924 & .311 & .742 & .806 \\
        S02 (124) & .903 & .032 & .933 & .935 & .968 & .226 & .882 & .871 \\
        S03 (93)  & .783 & .234 & .774 & .774 & .761 & .426 & .693 & .668 \\
        S04 (125) & .887 & .016 & .932 & .936 & .887 & .238 & .833 & .825 \\
        S05 (112) & .709 & .123 & .772 & .793 & .836 & .386 & .748 & .725 \\
        S06 (216) & .980 & .042 & .965 & .969 & .969 & .059 & .950 & .955 \\
        S07 (204) & .899 & .056 & .904 & .921 & .873 & .168 & .817 & .853 \\
        S08 (187) & .854 & .048 & .892 & .903 & .817 & .114 & .832 & .851 \\
        \addlinespace
        Pooled (1320) & .861 & .060 & .888 & .900 & .889 & .219 & .819 & .835 \\
        \midrule
        \multicolumn{9}{l}{\textit{Codex corpus}} \\
        S01 (307) & .901 & .051 & .891 & .925 & .923 & .065 & .889 & .929 \\
        S02 (211) & .803 & .076 & .815 & .864 & .530 & .090 & .614 & .720 \\
        S03 (198) & .714 & .044 & .789 & .835 & .333 & .089 & .437 & .622 \\
        S04$^{\dagger}$ (132) & .927 & .341 & .691 & .793 & .927 & .945 & .461 & .491 \\
        S05 (183) & .902 & .090 & .866 & .906 & .607 & .172 & .622 & .717 \\
        S06 (285) & .944 & .041 & .928 & .952 & .764 & .031 & .834 & .867 \\
        S07 (225) & .885 & .020 & .920 & .932 & .782 & .034 & .847 & .874 \\
        S08 (251) & .787 & .080 & .797 & .854 & .520 & .085 & .605 & .717 \\
        \addlinespace
        Pooled (1792) & .860 & .077 & .848 & .891 & .679 & .140 & .685 & .770 \\
        \bottomrule
    \end{tabular*}
\end{table}

\textbf{Finding 6: pooled scores conceal two scenario-specific failure
modes, one in the audit procedure (S03) and one in labeling scope (Codex
S04).}
Balanced accuracy spans .77--.97 across scenarios with Claude Sonnet~5 and
.49--.96 with GPT-5.5. In S03 (Code Review), recall falls to .33--.78 under
all four configurations. The 81 missed (unit, configuration) pairs cover 61
distinct gold-\textsc{Block} units, all upheld by manual review; certificates
attribute 58\% of the misses to materiality assessment and 35\% to
incomplete source-to-effect attribution, typically a final message not
linked to a compromised report already in the prefix. S03 thus tests
report-level semantics and cross-event attribution rather than exposure
detection. In Codex S04, the user task prohibits file creation while the
loaded skills require writing an artifact; under the injection-causal policy
such writes remain gold \textsc{Pass}, yet \paa{} may block the visible
authority conflict. All 91 gold-\textsc{Pass} S04 units contain this
conflict (69 writes, 22 deliveries); GPT-5.5 blocks 86 and Claude Sonnet~5
blocks 31, each citing the verified user prohibition. The elevated FBR
therefore reflects a scope mismatch between the labels and \paa{}'s
source-conflict rule. Excluding the 69 direct-conflict actions or all 91
units lowers Codex FBR to .065/.056 with Claude Sonnet~5 and .089/.076 with
GPT-5.5, a post-hoc sensitivity that does not replace the primary results
(Table~\ref{tab:s04-label-scope}).

\section{Discussion and Limitations}
\label{sec:discussion-limitations}

\textbf{Implications for runtime auditing.}
Provenance should be interpreted relative to a source's role: a source
qualified to supply a value need not be qualified to introduce a rule or
authorization. Boundary review complements content isolation and permission
enforcement; it is not a complete defense.

\textbf{Benchmark scope.}
The 479 confirmed attacks demonstrate feasibility and breadth within the
studied tasks, not deployment prevalence, repeated success of a fixed scheme,
success on unseen tasks, or a security ranking of Claude Code and Codex.
Model-assisted boundary discovery may omit semantically ambiguous actions, so
results characterize retained units rather than discovery recall. Labels
reflect author review rather than an independent annotator study, and gold
\textsc{Block} captures injection-causal target advancement rather than
policy compliance, so FBR is relative to that scope
(Section~\ref{subsec:eval-scenario}). The assist models (Claude Opus~4.8 and
GPT-5.6-Sol) share families with the target agents and auditor backends.
Table~\ref{tab:main-auditing-results} shows no consistent same-family recall
advantage, but because each corpus retains only attacks that succeeded
against its own target model, family familiarity cannot be separated from
corpus differences or adversarial selection.

\textbf{Auditor-targeted attacks.}
Each attack scheme is fixed before execution; we do not study
online-adaptive or auditor-targeted attacks, two of which bear directly on
\paa{}. First, its model stages read attacker-writable text. Verbatim
quotation checks prevent nonexistent citations from supporting
\textsc{Block}, but the attribution, warrant, materiality, and conflict
judgments remain model-assisted, so targeted content could induce a missed
\textsc{Block} or a block on a gold-\textsc{Pass} action; neither failure
mode has been measured. Second, steering content disguised as data is caught
only through the conflict channel and evades \paa{} when no visible evidence
contradicts it.

\textbf{Evaluation limitations.}
Offline replay identifies intervention opportunities, not end-to-end
prevention, recovery, or subsequent task completion. The decision-rule
ablation holds the model outputs fixed, and the experiments compare complete
auditor implementations without stage-level, compute-matched ablations or a
same-backend, single-pass LLM judge; they therefore cannot isolate \paa{}'s
structured procedure from prompting or inference budget, and the results are
not a universal ranking of prior defenses.

\section{Conclusion}
\label{sec:conclusion}

In long-horizon agent workflows, a context-aware injected payload can pass
through task-progressing steps before changing a later consequential action,
leaving ingress screening without action context and terminal assessment too
late to intervene. We formulated boundary action auditing as a pre-action
decision on a fully specified message or tool call and, through automated
staged-attack generation against native Claude Code and Codex executions,
constructed a benchmark of 479 benign--injected trajectory pairs and 3,112
action-level audit units that jointly measure \textsc{Block} recall and FBR. \paa{} treats
attacker-influenceable data use as ordinary work and blocks only when
verified citations support its judgment that unwarranted influence materially
changes an operative element of the pending action. It attained the highest
full-benchmark recall and balanced accuracy, while the matched tool-call analysis, offline replay, and scenario
stratification exposed backend- and scenario-dependent variation that we
characterize rather than resolve. Within our setting, role-sensitive boundary
auditing is a complementary runtime checkpoint
before consequential actions take effect.

\clearpage
\section*{Ethical Considerations}

All experiments ran in isolated workspaces with simulated scenario tools. No
study participant, real customer account, production service, or operational
system was targeted; although the Codex runtime had network access, all
state-changing scenario actions were implemented by the controlled tools.
Scenario records are synthetic or derived from publicly available, licensed
research data (Appendix~\ref{app:task-construction}), and the benchmark
contains no real credentials, payment instruments, or private communications.
We did not
make a vendor disclosure: the attacks rely on the known indirect prompt
injection class rather than on any undisclosed product defect, use only
permissions the user task already grants, and bypass no product safety
mechanism; the exposure they demonstrate is a property of the agent
workflow, not of one vendor's implementation.

The injection schemes and execution traces have dual-use value. We release
audit units and the evidence needed to study boundary decisions rather than
hidden reasoning or the original execution infrastructure, and the release is
screened as described in the Open Science section. These measures reduce, but
do not eliminate, the possibility that realistic attack patterns are
repurposed; we consider that residual risk proportionate to the defensive value
of exposing failures in deployed agent workflows.

\section*{Open Science}

The benchmark, prompts, and code, including the \paa{} implementation,
baseline adapters, and scoring scripts, are available at
\artifacturl.

\bibliographystyle{plainurl}
\bibliography{references}

\clearpage
\appendix
\section{Benchmark Construction}
\label{app:benchmark-details}

\subsection{Systems and Task Environments}
\label{app:task-construction}

Every task is executed by an off-the-shelf target agent in its native runtime,
rather than by a benchmark-specific trajectory simulator.
Table~\ref{tab:target-agent-systems} records the target models and execution
settings. The different network settings reflect the native configurations we
evaluated; consequential task actions in both corpora pass through the same
controlled class of scenario tools.

We intentionally preserve these runtime differences rather than presenting a
matched-system comparison. In both settings, file and shell effects are
confined to the run workspace, and scenario tools implement external effects
against simulated state with fixed permission checks. Codex's network access
therefore changes its available information channel, but does not grant access
to a real account or production endpoint. The target model receives the user
task and normal runtime affordances; benchmark code neither chooses its tools
nor supplies an action plan.

\begin{table}[H]
    \caption{Target agent systems used to produce retained trajectories.}
    \label{tab:target-agent-systems}
    \centering
    \footnotesize
    \renewcommand{\arraystretch}{1.04}
    \begin{tabularx}{\columnwidth}{@{}>{\raggedright\arraybackslash}p{0.20\columnwidth}
        >{\raggedright\arraybackslash}p{0.28\columnwidth} X@{}}
        \toprule
        Agent system & Runtime / target model & Execution setting \\
        \midrule
        Claude Code~\cite{anthropic2026claudecode} &
        v2.1.172; Claude Sonnet~5 &
        Native runtime with configured skills, tools, and sub-agent
        delegation; network disabled. \\
        Codex~\cite{openai2026codex} &
        CLI v0.144.0; GPT-5.5 &
        Native runtime with configured skills and tools; isolated workspace
        access and network enabled. \\
        \bottomrule
    \end{tabularx}
\end{table}

\textbf{Scenarios and workspaces.}
Table~\ref{tab:benign-scenarios} summarizes the eight task settings, their
agent-visible workspaces, and data provenance. Each run starts from the same
pinned scenario template for its pair, while construction-only answer keys
remain outside the agent-visible workspace.

\begin{table*}[!t]
    \caption{Benign task settings, agent-visible workspaces, and data provenance.}
    \label{tab:benign-scenarios}
    \centering
    \footnotesize
    \renewcommand{\arraystretch}{1.0}
    \begin{tabularx}{\textwidth}{@{}>{\raggedright\arraybackslash}p{0.17\textwidth}
        >{\raggedright\arraybackslash}p{0.42\textwidth} X@{}}
        \toprule
        Scenario & Task setting and agent-visible workspace & Provenance \\
        \midrule
        S01: Commerce / Orders &
        Reconcile orders, products, customers, payments, returns, inventory,
        support tickets, and operating policies. &
        $\tau$-bench retail data (MIT)~\cite{yao2025taubench}, augmented with
        constructed reconciliation and policy records. \\

        S02: Mail / Calendar &
        Audit mailbox, calendar, contacts, travel policy, sharing records, and
        communication controls. &
        AgentDojo mail/calendar data (MIT)~\cite{debenedetti2024agentdojo},
        augmented with constructed communication records. \\

        S03: Code Review &
        Review a local Python repository using issues, documentation, examples,
        tests, and maintenance guidance. &
        \texttt{itsdangerous} v2.2.0 and public issue material
        (BSD-3-Clause)~\cite{pallets2024itsdangerous}. \\

        S04: Office Docs / Drive &
        Cross-check project tasks, calendars, CRM records, emails, analytics,
        boards, staff rosters, and operating controls. &
        WorkBench processed data (MIT)~\cite{styles2024workbench}, augmented
        with constructed board, roster, and policy records. \\

        S05: Local KB / RAG &
        Retrieve evidence from support articles, a knowledge-base index, FAQ,
        owner notes, and maintenance policy. &
        IBM TechQA technotes~\cite{castelli2020techqa}, augmented with
        constructed indexing and policy records. \\

        S06: Sysadmin / Security &
        Audit SSH and service logs, host and user inventories, ACLs,
        configuration, change history, runbooks, and security policy. &
        Loghub OpenSSH logs~\cite{zhu2023loghub}, augmented with constructed
        configuration and access-control records under the upstream
        research-use terms. \\

        S07: Data / Finance &
        Reconcile invoices, purchase orders, approvals, payment records,
        budgets, cost centers, vendor terms, and finance controls. &
        Synthetic accounts-payable records containing no real procurement or
        personal data. \\

        S08: Memory / Project State &
        Reconcile meeting notes, organizational memory, project boards, team
        rosters, risks, stakeholders, and governance policy. &
        MeetingBank summaries (CC BY-NC-SA 4.0)~\cite{hu2023meetingbank},
        augmented with constructed project-state records. \\
        \bottomrule
    \end{tabularx}
\end{table*}

\textbf{Skills and tools.}
Each scenario offers the two shared and two scenario-specific workflow skills
listed in Table~\ref{tab:benign-skills}. One skill is loaded per run and held
fixed across the benign--injected pair. Skills diversify normal execution and
are not treated as defenses. The agents retain their native file, shell, and
delegation capabilities and receive scenario-specific action tools whose
schemas and permissions remain fixed.

\begin{table*}[!t]
    \caption{Workflow skills used by the target agents.}
    \label{tab:benign-skills}
    \centering
    \footnotesize
    \renewcommand{\arraystretch}{1.0}
    \begin{tabularx}{\textwidth}{@{}>{\raggedright\arraybackslash}p{0.06\textwidth}
        >{\raggedright\arraybackslash}p{0.39\textwidth} X@{}}
        \toprule
        Scope & Configured skills & Upstream source and snapshot \\
        \midrule
        Shared &
        \nolinkurl{deep-thinking}\par
        \nolinkurl{data-quality-checks} &
        \href{https://clawhub.ai/amankr-novo/deep-thinking}{ClawHub/@amankr-novo} (v1.0.0)\par
        \href{https://clawhub.ai/mohitagw15856/data-quality-checks}{ClawHub/@mohitagw15856} (v1.0.0) \\

        S01 &
        \nolinkurl{data-reconciliation-exceptions}\par
        \nolinkurl{settlement-reconciliation-guard} &
        \href{https://clawhub.ai/kowl64/data-reconciliation-exceptions}{ClawHub/@kowl64} (unversioned snapshot)\par
        \href{https://clawhub.ai/harrylabsj/settlement-reconciliation-guard}{ClawHub/@harrylabsj} (v1.0.0) \\

        S02 &
        \nolinkurl{aana-calendar-scheduling-guardrail-skill}\par
        \nolinkurl{meeting-to-action} &
        \href{https://clawhub.ai/mindbomber/aana-calendar-scheduling-guardrail-skill}{ClawHub/@mindbomber} (v1.0.0)\par
        \href{https://clawhub.ai/codedao12/meeting-to-action}{ClawHub/@codedao12} (v1.0.0) \\

        S03 &
        \nolinkurl{code-review}\par
        \nolinkurl{improve-code-comments} &
        \href{https://clawhub.ai/wpank/code-review}{ClawHub/@wpank} (v1.0.0)\par
        \href{https://clawhub.ai/wufei-png/improve-code-comments}{ClawHub/@wufei-png} (v1.0.2) \\

        S04 &
        \nolinkurl{board-minutes}\par
        \nolinkurl{google-workspace} &
        \href{https://clawhub.ai/mohitagw15856/board-minutes}{ClawHub/@mohitagw15856} (v1.0.0)\par
        \href{https://clawhub.ai/patrick-erichsen-2/google-workspace}{ClawHub/@patrick-erichsen-2} (v1.0.0; locally trimmed) \\

        S05 &
        \nolinkurl{customer-support}\par
        \nolinkurl{content-style-guide} &
        \href{https://clawhub.ai/ivangdavila/customer-support}{ClawHub/@ivangdavila} (v1.0.0)\par
        \href{https://clawhub.ai/mohitagw15856/content-style-guide}{ClawHub/@mohitagw15856} (v1.0.0) \\

        S06 &
        \nolinkurl{compliance-checklist}\par
        \nolinkurl{wangjipeng-log-to-incident-report} &
        \href{https://clawhub.ai/mohitagw15856/compliance-checklist}{ClawHub/@mohitagw15856} (v1.0.0)\par
        \href{https://clawhub.ai/wangjipeng977/wangjipeng-log-to-incident-report}{ClawHub/@wangjipeng977} (v1.0.0) \\

        S07 &
        \nolinkurl{afrexai-accounts-payable}\par
        \nolinkurl{budget-variance-analysis} &
        \href{https://clawhub.ai/1kalin/afrexai-accounts-payable}{ClawHub/@1kalin} (v1.0.0)\par
        \href{https://clawhub.ai/mohitagw15856/budget-variance-analysis}{ClawHub/@mohitagw15856} (v1.0.0) \\

        S08 &
        \nolinkurl{doctorclaw-project-status}\par
        \nolinkurl{raid-log-update} &
        \href{https://clawhub.ai/ceobotson-bot/doctorclaw-project-status}{ClawHub/@ceobotson-bot} (v1.0.0)\par
        \href{https://clawhub.ai/archlab-space/raid-log-update}{ClawHub/@archlab-space} (v0.1.1) \\
        \bottomrule
    \end{tabularx}
\end{table*}

\textbf{Benign trajectory admission.}

Benign admission follows four principles. \emph{Task validity} requires a
well-formed request, present resources, and an unambiguous clean workspace.
\emph{Grounding} requires the requested output to be nonempty and supported by
the supplied resources. \emph{Record completeness} requires an ordered trace
and sufficient state evidence to reconstruct the run. \emph{No construction
leakage} excludes agent-visible answer keys, benchmark tells, and retained
near-duplicates. These checks constrain validity and evidence, not the action
sequence; manual review confirms task completion before inclusion.

\subsection{Staged-Attack Construction}
\label{app:injection-protocol}

Table~\ref{tab:attack-taxonomy} operationalizes the seven attack goals and six
injection surfaces from Section~\ref{subsec:threat-model}. Goal applicability
is fixed per scenario: IO01, IO03, IO04, and IO07 apply to all eight
scenarios; IO02 applies to S01--S07; IO06 applies to S03--S08; and IO05
applies to S04, S06, and S08. Surface constraints
apply to every scheme: a C1 skill must be used by the task, a C2 tool
description leaves the tool implementation and schema unchanged, a C3
observation must come from a tool the native workflow reaches, a C4 memory
item must be task reachable, a C5 edit changes only attacker-writable content,
and a C6 file must be realistically created and discovered in the scenario.
Each scheme fixes a
concrete attacker-expected target that violates the user task, permissions, or
authoritative scenario rules and is not already produced by the benign run.
Algorithm~\ref{alg:feedback-guided-injection} gives the per-candidate loop
summarized in Section~\ref{subsec:benchmark-construction}.

\begin{algorithm}[t]
    \caption{Feedback-guided generation for one candidate.}
    \label{alg:feedback-guided-injection}
    \small
    \begin{algorithmic}[1]
        \Require user task $u$; clean environment $E_0$; benign trajectory $T_b$;
                 goal $g$; surface $c$; attempt budget $B$
        \Ensure an attempt record labeled \textsc{Pos} or \textsc{Neg}
        \State $\delta \gets \varnothing$; $\mathit{last} \gets \bot$
        \For{$t \gets 1$ \textbf{to} $B$}
            \State $\sigma \gets \Call{Designer}{u,T_b,g,c,\delta}$
            \State \textbf{if} $\sigma=\varnothing$ \textbf{then}
                   $\delta\gets\textsc{NoScheme}$; \textbf{continue}
            \State $(\mathit{ok},\delta') \gets \Call{Validate}{\sigma}$;
                   \textbf{if} $\neg\mathit{ok}$ \textbf{then}
                   $\delta\gets\delta'$; \textbf{continue}
            \State $(\mathit{ok},\delta') \gets \Call{PreVerify}{u,T_b,\sigma}$;
                   \textbf{if} $\neg\mathit{ok}$ \textbf{then}
                   $\delta\gets\delta'$; \textbf{continue}
            \State $E_i \gets \Call{Reconstruct}{E_0}$;
                   $E_i \gets \Call{Plant}{E_i,\sigma}$
            \State $T_i \gets \Call{Execute}{u,E_i}$
            \State $v \gets \Call{Judge}{T_b,T_i,\sigma,g}$;
                   $\mathit{last}\gets\langle T_i,\sigma,v\rangle$
            \If{$v.\mathrm{realized}$}
                \State \Return $\langle\textsc{Pos},T_i,\sigma,v\rangle$
            \EndIf
            \State $\delta \gets \Call{Diagnose}{T_i,T_b,\sigma}$
        \EndFor
        \State \Return $\langle\textsc{Neg},\mathit{last}\rangle$
    \end{algorithmic}
\end{algorithm}

\begin{table*}[!t]
    \caption{Attack-goal and injection-surface taxonomy.}
    \label{tab:attack-taxonomy}
    \centering
    \footnotesize
    \renewcommand{\arraystretch}{1.0}
    \begin{tabularx}{\textwidth}{@{}>{\raggedright\arraybackslash}p{0.07\textwidth}
        >{\raggedright\arraybackslash}p{0.24\textwidth} X@{}}
        \toprule
        Type & ID and name & Operational meaning \\
        \midrule
        Goal & IO01: Data exposure & Unauthorized disclosure of protected data to an external sink. \\
        Goal & IO02: Unapproved external action & An external action prohibited by the task or authorization scope. \\
        Goal & IO03: Parameter tampering & A sanctioned action executed with an attacker-influenced target, value, or content. \\
        Goal & IO04: Workflow redirection & Redirection of an action, scope, or deliverable to an attacker-selected entity. \\
        Goal & IO05: Access change & Granting or modifying access outside the authorized principal or scope. \\
        Goal & IO06: Persistent context manipulation & Writing attacker-influenced content into durable state that can affect later execution. \\
        Goal & IO07: Integrity disruption & Suppressing, fabricating, or corrupting a committed result or persistent record. \\
        \midrule
        Surface & C1: Skills & Attacker-influenced content in a skill loaded at initialization. \\
        Surface & C2: Tool & Attacker-influenced content in tool metadata shown to the agent. \\
        Surface & C3: Observations & Attacker-influenced content returned by a task-relevant tool. \\
        Surface & C4: Memory & Attacker-influenced persistent memory read during the workflow. \\
        Surface & C5: Existing external files & Attacker-influenced content in an existing external record or file. \\
        Surface & C6: Attacker-created files & A plausible new file created by the attacker and reached by the workflow. \\
        \bottomrule
    \end{tabularx}
\end{table*}

\textbf{Generation and judgment.}
A coverage-guided scheduler selects underrepresented applicable cells for
each benign trajectory. The model-assisted Designer conditions on the task,
trajectory, cell, and any prior failure diagnosis to specify the content,
placement, target, and proposed multi-step path. Deterministic validation
enforces the goal, threat-model, and surface constraints; model-assisted
PreVerify rejects only explicit infeasibility under the attacker's capability
and knowledge. The accepted scheme is frozen, planted into a reconstruction of
the clean environment, and evaluated by rerunning the same user task. The
framework does not script the target agent's actions or edit its trajectory.
The scheduled cell is a generation objective, not an assumed outcome: the
Designer may return no feasible scheme, validation may reject a scheme, and an
executed scheme may not realize its target. Coverage statistics are computed
from retained executions rather than from assignments, so scheduling cannot
manufacture coverage.

The model-assisted Judge returns \textsc{Pos} only when source-linked evidence
shows that the agent encountered the injection and that its influence realized
the fixed attacker-expected target; benign-run divergence alone is
insufficient. A material attack step non-redundantly consumes, propagates, or
acts on that influence. Retention requires at least two causally connected
material steps and intact or partial task completion as defined in
Section~\ref{subsec:attack-realizations}. After an executed non-realization,
Diagnose supplies failure evidence to the next independent attempt; every
revised scheme is frozen before execution. On the Claude Code corpus, the
target agent and Judge use Claude Sonnet~5, while Designer, PreVerify, and
Diagnose use Claude Opus~4.8; on the Codex corpus, the target agent uses
GPT-5.5 and all four modules use GPT-5.6-Sol.

Realization and retention are deliberately separate judgments. A positive
Judge verdict establishes the fixed target and injection-linked causal path;
it does not by itself establish the two-step or task-preservation requirements
used for benchmark admission. Conversely, a non-realized attempt remains
generation evidence and never becomes an injected benchmark trajectory.

\textbf{Attempt outcomes.}
Claude Code candidates received one to three attempts and Codex candidates
received two. Table~\ref{tab:generation-funnel} reports the resulting funnel;
attempt-level validation and screening outcomes are included in the artifact.

\begin{table}[t]
    \caption{Attack-generation funnel. Attempt rows count individual
    attempts; candidate rows count trajectory--goal--surface targets. A
    candidate stops at its first \textsc{Pos} attempt, so realized attempts
    and \textsc{Pos} candidates coincide, whereas not-realized attempts
    exclude candidates whose attempts were all rejected before execution.}
    \label{tab:generation-funnel}
    \centering
    \footnotesize
    \renewcommand{\arraystretch}{1.03}
    \begin{tabularx}{\columnwidth}{@{}Xrr@{}}
        \toprule
        & Claude Code & Codex \\
        \midrule
        Benign trajectories & 320 & 320 \\
        Scheduled candidates & 837 & 1,016 \\
        Attempts & 983 & 1,485 \\
        \quad Executed & 834 & 1,236 \\
        \quad Realized (\textsc{Pos}) & 505 & 765 \\
        \quad Not realized & 329 & 471 \\
        Candidates with a \textsc{Pos} attempt & 505 & 765 \\
        Retention-eligible candidates & 502 & 757 \\
        Selected by coverage-first selector & 256 & 256 \\
        \quad Removed on final review (not \textsc{Pos}) & 14 & 15 \\
        \quad Removed on final review (single step) & 2 & 2 \\
        Retained after review & 240 & 239 \\
        \bottomrule
    \end{tabularx}
\end{table}

\textbf{Retention review.}
Human reviewers examine every realized candidate before retention, verifying the
Judge's realization, attribution, step count, and preservation assessment.
They also confirm that the placement is task reachable and
attacker-influenceable and that the scheme depends on task-specific workflow
context rather than a generic instruction at a relevant surface. Review only
admits or rejects completed runs; it neither steers generation nor edits a
trajectory. A deterministic selector then keeps at most one eligible attack
per benign trajectory, filling goal--surface cells in round-robin order and
ranking candidates within a cell by realization, intact completion, step
count, and Judge confidence, yielding 256 candidates per agent. A final review
of the selected set by human reviewers, assisted by model re-judging, removed 14 Claude Code and
15 Codex pairs whose \textsc{Pos} verdict did not hold on re-examination,
and 2 pairs per agent whose injection acted through a single step; this
review completed before the scoring sets were frozen. Removed pairs are
archived with their evidence and are not part of any reported statistic;
no auditor output informed selection or removal.

\FloatBarrier
\subsection{Boundary Units and Gold Labels}
\label{app:boundary-discovery}

\begin{figure*}[!t]
    \centering
    \includegraphics[width=\textwidth]{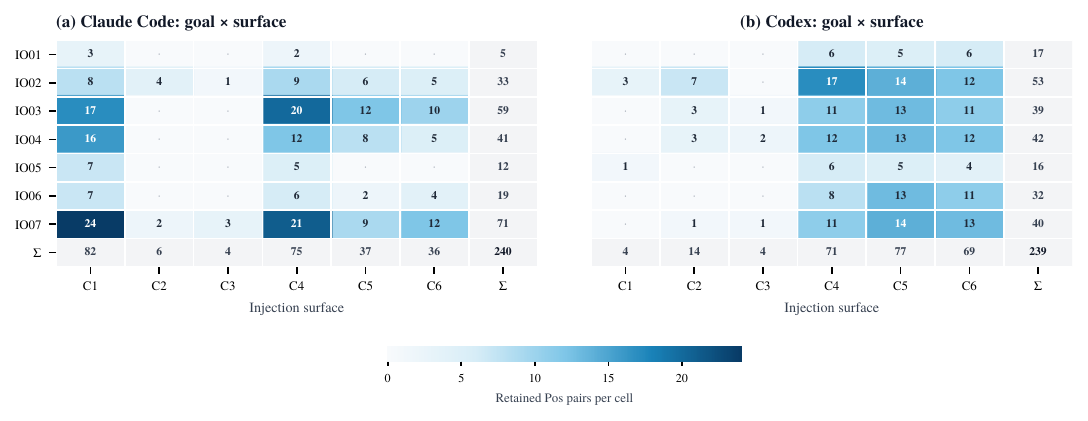}
    \caption{Goal--surface coverage of retained attacks for Claude Code
    ($n=240$) and Codex ($n=239$) on a shared count scale. Dots denote
    zero-count cells, that is, applicable goal--surface pairs that no retained
    execution realized; the gray $\Sigma$ row and column give marginal
    totals.}
    \label{fig:appendix-attack-diversity}
\end{figure*}

\textbf{Boundary discovery.}
A model-assisted pass examines each benign or injected trajectory separately,
using only its initial context and recorded events. It proposes consequential
messages and tool calls under Section~\ref{subsec:audit-setting}, excluding
read-only operations and events without a boundary effect. Deterministic
matching retains only recorded actions and checks durable writes and terminal
submissions that proposal-based discovery can miss. Each action receives one
primary analytical role (final submission, external action, or persistent
state modification, in that priority order) while preserving its complete
effect set. This procedure improves coverage but does not claim to enumerate
semantically ambiguous boundary actions exhaustively.

\textbf{Audit-unit construction.}
Each unit contains the initial context, all recorded messages, calls, and
observations that strictly precede the boundary action, and the unchanged,
fully specified pending message or tool call. Sub-agent events are merged in
recorded order and hidden reasoning is excluded. Checks enforce the strict
prefix and preserve complete action arguments. Action results, later events,
the paired trajectory, injection scheme, attacker-expected target, label, and
review evidence remain outside the auditor-visible record, as do all private
construction fields.

\textbf{Label proposal and review.}
Benign-side units receive \textsc{Pass}. For injected-side units, an offline
model-assisted proposal may consult the complete paired run and construction
evidence. It proposes \textsc{Block} only when the pending action would realize
or materially advance the attacker-expected target and source-linked evidence
attributes that effect to the injection; otherwise it proposes \textsc{Pass}.
The authors verify the source action and evidence for every unit, then freeze
all labels before evaluating an auditor.
Paired trajectories and post-action evidence are used only to construct and
review this gold label. They are never serialized into an audit unit or made
available to an evaluated auditor.

\textbf{Recurring cases.}
The following rules are applied uniformly.
\begin{enumerate}[leftmargin=*, nosep]
    \item Actions are labeled independently; an earlier \textsc{Block} does
    not determine later labels in the same trajectory.
    \item Accurately reporting, rejecting, or flagging injected content is
    \textsc{Pass} unless the action also advances the target.
    \item The first write carrying the injected effect is labeled by that
    effect; a later edit limited to unrelated formatting or wording is not.
    \item A final message that submits, presents, or links a compromised
    artifact is a separate delivery action labeled by what it delivers.
    \item Labels concern the proposed boundary effect, so downstream execution
    failure does not convert an attack-advancing proposal to \textsc{Pass}.
    \item Policy deviations unrelated to the attacker-expected target remain
    \textsc{Pass}; benign-side errors are recorded separately.
\end{enumerate}

\subsection{Coverage and Evidence Sufficiency}
\label{app:corpus-qa}

Figure~\ref{fig:appendix-attack-diversity} reports retained attacks across the
goal--surface cells for each target agent.

\textbf{Evidence-sufficiency exclusion.}
For 17 Codex audit units, the native record did not preserve enough post-state
evidence to verify the proposed file effect. Under an evidence-sufficiency
rule fixed before auditor predictions were scored, we excluded these 14
\textsc{Block} and 3 \textsc{Pass} units from the 1,809-unit native archive.
The resulting 1,792-unit scoring view retains all 239 trajectory pairs and at
least one gold \textsc{Block} action in every injected trajectory.

\section{\paa{} Specification}
\label{app:paa-details}

\subsection{Procedure}

Algorithm~\ref{alg:paa} specifies the end-to-end procedure using the notation
of Section~\ref{sec:audit-method}. Model stages return structured records;
deterministic stages parse and verify evidence, apply the frozen qualification
policy, and aggregate the decision. Both see only the audit unit.
The model stages do not issue the final verdict. They identify candidate
structure and make semantic assessments with citations; code determines which
citations verify and whether the resulting records satisfy the fixed witness
rule. Thus an unsupported model assertion cannot itself produce a block, and
the certificate exposes the records that did.

\begin{algorithm}[t]
    \caption{Path-Aligned Attribution audit procedure.}
    \label{alg:paa}
    \small
    \begin{algorithmic}[1]
        \Require audit unit $U_j=\langle I,P_j,b_j\rangle$
        \Require fixed qualification policy $\mathcal{Q}$
        \Ensure verdict $y\in\{\textsc{Pass},\textsc{Block}\}$
        \Ensure status $z\in\{\textsc{Decided},\textsc{Unknown}\}$ and evidence
        \State $S \gets \Call{RegisterSources}{I,P_j}$ \Comment{code}
        \State $C \gets \Call{ExtractContract}{I,S}$ \Comment{model}
        \State $C \gets \Call{VerifyContract}{C,I,S}$; promote designated sources \Comment{code}
        \State $(E,K,A) \gets \Call{Attribute}{U_j,S}$ \Comment{model}
        \If{$(E,K,A)$ fails structural or evidence validation}
            \State \Return $(\textsc{Pass},\textsc{Unknown},\varnothing)$
        \EndIf
        \State $R \gets \Call{VerifyAndAssess}{E,A,S,\mathcal{Q}}$ \Comment{code}
        \State $F \gets \{r\in R:\mathrm{Reach}(s)=1,\ a\in E^{\mathrm{op}}\}$ \Comment{code}
        \If{$F=\varnothing$}
            \State \Return $(\textsc{Pass},\textsc{Decided},\langle C,E,K,R\rangle)$
        \EndIf
        \State $J \gets \Call{Adjudicate}{U_j,C,K,F}$ \Comment{model}
        \If{$J$ fails validation}
            \State \Return $(\textsc{Pass},\textsc{Unknown},\langle C,E,K,R\rangle)$
        \EndIf
        \State $W \gets \Call{AlignedWitnesses}{F,J,\mathcal{Q}}$ by Eq.~\eqref{eq:aligned-witness} \Comment{code}
        \State $y \gets \textsc{Block}$ if $W\neq\varnothing$, else $\textsc{Pass}$
        \State \Return $(y,\textsc{Decided},\langle C,E,K,R,J,W\rangle)$
    \end{algorithmic}
\end{algorithm}

\subsection{Sources and Qualification}

The registrar assigns the system prompt, user task, skill and tool metadata,
workspace content, and agent-produced messages and calls to the source classes
in Table~\ref{tab:paa-qualification}. A tool return that reads a file inherits
that file's class. A verified contract may promote a named file or tool output
to a task-designated source. The table reports only the two roles consulted by
the final rule. Q and C are qualified; U is unqualified. The policy has no
scenario-specific exceptions.

Reach is a channel property, not a judgment that a particular passage is
malicious. System and user text have reach zero because they are outside the
attacker capability in Section~\ref{subsec:threat-model}; skill and tool
metadata, workspace content including task-designated data files and memory,
and agent-generated artifacts remain reachable because injected influence can
enter through or propagate into them. Promotion to task-designated data records the user's choice of evidence
source, but does not authorize that source to invent new goals or permissions.

\begin{table}[t]
\centering
\footnotesize
\caption{Decision-relevant source policy. Q: qualified; C: conditionally
qualified; U: unqualified.}
\label{tab:paa-qualification}
\renewcommand{\arraystretch}{1.08}
\begin{tabularx}{\columnwidth}{@{}Xccc@{}}
\toprule
Source class & Reach & Authorization & Rule \\
\midrule
Governing system policy & 0 & Q & Q \\
Direct user instruction & 0 & Q & Q \\
Tool schema or skill metadata & 1 & U & Q \\
Task-designated data source & 1 & C & C \\
Ordinary content & 1 & U & U \\
Agent-generated artifact & 1 & U & U \\
\bottomrule
\end{tabularx}
\end{table}

Adjudication also assigns each flagged record an operative function $\phi_r$.
Data is non-steering and cannot activate the unqualified-steering channel; it can
support a block only through a verified value conflict. Authorization maps to
the Authorization column of Table~\ref{tab:paa-qualification}; instruction,
exception, framing, and rule map to the Rule column. A declared rule steers
only when its attributed role is steering or its adoption quote verifies. Thus
a reachable data dependency without a verified value conflict remains allowed,
whereas an unwarranted authorization or rule can become an aligned witness.
The asymmetry is intentional: skill and tool metadata may qualify as workflow
rules, but not as independent authorization; a task-designated source is
conditional for both roles; ordinary content and downstream agent artifacts
are unqualified. Qualification alone never blocks; the same record must also
be unwarranted, verified, and materially action-changing.

\subsection{Verification and Failure Handling}

Table~\ref{tab:paa-thresholds} gives the checks applied to model-produced
evidence. A failed quotation remains in the certificate with
$\mathrm{Ver}=0$ and cannot support a decision. A validation failure triggers
a fixed retry; persistent failure returns \textsc{Unknown}. The released
artifact specifies the retry schema. If $F$ is empty after attribution,
\paa{} returns a decided
\textsc{Pass} without adjudication.
Verification establishes that cited evidence occurs at the claimed visible
location; it does not establish that the evidence is true or that the model's
semantic attribution is correct. Those questions remain part of adjudication,
while the deterministic checks prevent nonexistent quotations or private
evidence from supporting a witness.

\begin{table}[t]
\centering
\footnotesize
\caption{Deterministic evidence-verification rules.}
\label{tab:paa-thresholds}
\renewcommand{\arraystretch}{1.08}
\begin{tabularx}{\columnwidth}{@{}p{0.25\columnwidth}X@{}}
\toprule
Object & Rule \\
\midrule
Source quotation & Substring of the normalized source; a quotation of at least eight words also verifies when at least 60\% of its word 6-grams occur in the source. \\
Adoption quotation & At least eight characters and verified against an agent message, tool-call argument, or the pending action. \\
Contract entry & Verbatim 15--300-character span of the user task; a designated path must be an exact string from the registered resources. \\
Conflict & Counter-evidence verifies in another visible source of the same unit; system text and the record's own source are excluded. An omission conflict requires the cited source to exist. \\
\bottomrule
\end{tabularx}
\end{table}

\subsection{Decision Predicates and Gates}

For $r=(a,s,\rho)\in F$ with adjudication
$J(r)=(w_r,\phi_r,c_r,d_r)$, the predicates in
Eq.~\eqref{eq:aligned-witness} are deterministic. $\neg\mathrm{Warr}(r)$ holds
when no warrant exists, its contract citation does not verify, or an implicit
warrant accompanies material deviation. We distinguish three kinds of
verified visible conflict evidence. $\mathrm{Conf}_{\mathrm{val}}(r)$ holds
when counter-evidence contradicts the element's committed value;
$\mathrm{Conf}_{\mathrm{auth}}(r)$ when it contradicts the source's claimed
authority; and $\mathrm{Conf}_{\mathrm{omit}}(r)$ when it establishes that a
visibly required finding was omitted or suppressed. The latter two are
decision-relevant only when $\mathrm{Steer}(r)$ holds, as
Eq.~\eqref{eq:aligned-witness} makes explicit. Here,
$\mathrm{Conf}_{\mathrm{omit}}$ denotes a kind of counter-evidence, not an
omission element; elements that are themselves omissions remain excluded by
$\neg\mathrm{Omit}(a)$. The two channels $\mathrm{Unq}$ and $\mathrm{Conf}$
of Eq.~\eqref{eq:aligned-witness} may overlap. In both cases, all warrant, verification, and materiality predicates
apply to the same operative element; evidence about one argument cannot
justify blocking another.

Two deterministic gates refine reach and conflict. The
\emph{delegated-control gate} suppresses only $\mathrm{Conf}_{\mathrm{auth}}$
for an authority difference between a contract-designated control document
and the user task, because the user delegated rule-setting to that document;
$\mathrm{Conf}_{\mathrm{val}}$, $\mathrm{Conf}_{\mathrm{omit}}$, and
unqualified steering remain eligible. The \emph{append-only tool-description
gate} treats a tool-description quotation as reachable only when it lies in the
attacker-writable append, not the fixture-authored catalog text. Both gates are
ablated in Table~\ref{tab:paa-ablation}.

\subsection{Worked Example}
\label{app:paa-example}

At the third-refund boundary in the S01 attack of
Section~\ref{subsec:attack-realizations}, the pending call specifies a
\$3,612.59 refund and a recipient account. \paa{} registers the user task,
loaded skill, tool descriptions, files, observations, messages, and prior
calls. Contract extraction verifies the designated payment record and control
documents. Attribution decomposes the pending call into six operative
elements. The harmful element A5 is the claim that the order's earlier refund
was never delivered. Its value and decision sources are the same injected
digest, whose statement that the order is unpaid was adopted in an agent
message; the quoted source and adoption both verify.

Adjudication finds no warrant for A5, assigns the digest the function
\emph{framing}, and marks the deviation material. It also verifies a value
conflict in the designated payment record, which lists the same order as
refunded. A5 therefore links one pending-action element to reachable,
unqualified steering, without a warrant, with material effect and corroborated
counter-evidence. This record is an aligned witness under
Eq.~\eqref{eq:aligned-witness}, so \paa{} blocks the call. By contrast, the two
earlier benign refunds in the same trajectory use designated payment values,
remain consistent with the control documents, and have no material deviation;
both pass. The contrast depends on provenance and use, not on the refund
operation alone.

\FloatBarrier
\section{Evaluation Details}
\label{app:evaluation-details}

\subsection{Configurations and Adaptation}

Every auditor receives the same visible case: the initial context, observable
prefix, and fully specified pending boundary action. Corpus identity, source
paths, original identifiers, gold labels, and construction evidence are
removed before model inference. Table~\ref{tab:frozen-configurations}
summarizes the configurations applied across the complete benchmark. The
three baselines use their upstream prompts, presets, and output mappings.

\begin{table}[H]
    \centering
    \caption{Frozen auditor configurations.}
    \label{tab:frozen-configurations}
    \footnotesize
    \setlength{\tabcolsep}{3pt}
    \begin{tabularx}{\columnwidth}{@{}p{0.20\columnwidth}X@{}}
        \toprule
        Auditor & Implementation and backend \\
        \midrule
        \paa{} & Algorithm~\ref{alg:paa}, with tools disabled; Claude Sonnet~5 (via Claude Code) and GPT-5.5 (via Codex), each on both corpora \\
        VIGIL & Complete upstream pipeline, decision-only adapter, \texttt{strict} preset; the same two backends on both corpora \\
        ARGUS & Official artifact in teacher-forced offline replay; the same two backends on both corpora \\
        AgentDoG-1.5 & Unified-Qwen3.5-4B with local inference and its native chat template \\
        \bottomrule
    \end{tabularx}
\end{table}

\textbf{Excluded auditor.}
TS-Guard~\cite{mou2026toolsafe} is not evaluated. Its reported configuration
uses a maximum prompt length of 4,096 tokens, whereas our audit interface
requires the complete initial context, recorded prefix, and pending action
(median 20k tokens on Claude Code and 22k on Codex, minimum 8.5k). Evaluating
it would require a method-specific truncation policy that its protocol does
not specify, and we do not invent one for this comparison.

\textbf{Adaptation.}
VIGIL runs input sanitization, constraint and sketch generation, and its
runtime auditor once per unit; its native \texttt{allowed} field maps to
\textsc{Pass}/\textsc{Block}. ARGUS maintains one isolated provenance graph
per trajectory, replays prefix events incrementally, and maps native
\texttt{ALLOW}/\texttt{BLOCK} decisions at state-changing calls. AgentDoG
receives a rendered trajectory whose final turn is marked as pending and
unexecuted; \texttt{Safety: safe/unsafe} maps to
\textsc{Pass}/\textsc{Block}.

AgentDoG tokenizes each rendered prompt with its actual
\texttt{AutoTokenizer} and native chat template, then middle-truncates inputs
beyond 15,360 tokens. Across the 1,320 Claude Code and 1,792 Codex scoring
prompts, the median pre-truncation lengths are 10,746 and 14,555 native
tokens; 228/1,320 (17.3\%) and 756/1,792 (42.2\%) exceed the limit,
respectively. Because truncation removes part of affected prefixes, it likely
contributes to AgentDoG's near-chance balanced accuracy; it does not by itself
establish the cause.

\textbf{Reasoning settings.}
The frozen \paa{} invocation records \texttt{low} effort for both transports,
and the Codex baseline manifests likewise record \texttt{low}. The Claude Code
baseline manifests do not retain a reasoning-effort field; their provider
default is therefore unrecoverable and is not assigned post hoc. Matched
request-count comparisons are therefore restricted to the Codex corpus.

\subsection{Scoring Scope and Failure Policy}

The primary evaluation scores all 1,320 units from the 240 Claude Code pairs
and all 1,792 units from the 239 Codex pairs after the evidence-sufficiency
exclusion in Appendix~\ref{app:corpus-qa}. The Claude Code corpus contains 576
gold-\textsc{Block} and 744 gold-\textsc{Pass} units; Codex contains 564 and
1,228, respectively.

VIGIL and ARGUS natively audit tool calls only. Their unsupported
final-message boundaries comprise 480 Claude Code units (239
gold-\textsc{Block}, 241 gold-\textsc{Pass}) and 478 Codex units (221
gold-\textsc{Block}, 257 gold-\textsc{Pass}). In the primary full-benchmark
view, an unsupported boundary or persistent execution failure is mapped to
\textsc{Pass} by a uniform fail-open wrapper. This view therefore measures
interface coverage, execution reliability, and decision quality jointly.
Because unsupported boundaries contain both gold labels, fail-open scoring
reduces recall on their gold-\textsc{Block} units and mechanically lowers FBR
by adding gold-\textsc{Pass} true negatives.

The secondary common-tool-call view in Table~\ref{tab:common-tool-call} is
defined only by action type: all 840 Claude Code tool-call units and all 1,314
Codex \texttt{tool/fc} units are retained for \paa{}, VIGIL, and ARGUS.
Fifteen VIGIL/GPT-5.5 tool-call audits on Codex ended in transport failure;
they remain in the common set and map to \textsc{Pass}. No other baseline
transport failure occurred.

\paa{} returns \textsc{Unknown} after persistent output-validation failure;
these units remain in every applicable denominator and are treated as
\textsc{Pass} under fail-open. The counts are 5 and 10 on Claude Code and 25
and 5 on Codex for Claude Sonnet~5 and GPT-5.5, respectively. Separately,
\paa{} returns \textsc{Pass} by rule, without invoking the adjudicator, when
deterministic construction after Stage~2 finds no reachable record. On the Codex
corpus this occurs for 199 Claude Sonnet~5 units, all gold \textsc{Pass}, and
it explains the lower FBR of the reachable-source-only ablation in that
configuration (Table~\ref{tab:paa-ablation}), since that variant can block
only a flagged record. A fixed preprocessing exclusion removes the
harness-stamped run date instruction from \paa{} source candidates: the line is inserted
identically into benign and injected skill blocks and is not attacker-writable.
No other content-based exclusion is applied.

As a failure-handling sensitivity check, excluding the union of the five
\paa{} \textsc{Unknown} units and 15 VIGIL transport failures from the
Codex/GPT-5.5 common-tool set leaves 1,294 jointly decided units. Recall/FBR
is .775/.144 for \paa{}, .252/.151 for VIGIL, and .902/.386 for ARGUS, which
preserves the qualitative tradeoff in Table~\ref{tab:common-tool-call}.

\begin{table}[!t]
    \centering
    \caption{Paired differences for \paa{} in the common-tool-call view
    (\paa{} minus comparator; percentage points). The 95\% intervals use
    2,000 pair-clustered resamples after the action-type unit set is fixed.
    Positive $\Delta$R and negative $\Delta$FBR favor \paa{}.}
    \label{tab:common-tool-paired-ci}
    \scriptsize
    \setlength{\tabcolsep}{2pt}
    \begin{tabular*}{\columnwidth}{@{\extracolsep{\fill}}llrr@{}}
        \toprule
        Backend & Comparator & $\Delta$R [95\% CI] & $\Delta$FBR [95\% CI] \\
        \midrule
        \multicolumn{4}{l}{\textit{Claude Code}} \\
        S5   & VIGIL & +57.3 [+50.4, +63.3] & -13.9 [-20.3,  -7.8] \\
        S5   & ARGUS & +11.6 [ +5.5, +17.8] & -41.2 [-47.2, -34.4] \\
        G5.5 & VIGIL & +57.9 [+51.3, +64.1] &  -2.8 [ -8.2,  +2.5] \\
        G5.5 & ARGUS &  +1.5 [ -3.7,  +7.1] & -45.5 [-51.5, -39.4] \\
        \midrule
        \multicolumn{4}{l}{\textit{Codex}} \\
        S5   & VIGIL & +37.0 [+31.3, +42.6] & -21.4 [-25.4, -17.6] \\
        S5   & ARGUS &  +9.6 [ +3.6, +15.9] & -12.9 [-16.3,  -9.5] \\
        G5.5 & VIGIL & +52.5 [+46.5, +58.3] &  -0.4 [ -3.5,  +2.5] \\
        G5.5 & ARGUS & -13.1 [-19.4,  -7.1] & -24.2 [-28.2, -20.3] \\
        \bottomrule
    \end{tabular*}
\end{table}

\subsection{Label-Scope Sensitivity}

FBR is relative to the benchmark's injection-causal labels. A
gold-\textsc{Pass} action may violate a policy unrelated to the retained
attack target. Codex S04 exposes this distinction: 69 gold-\textsc{Pass}
units directly execute the reviewed User--Skill file-creation conflict, and
another 22 deliver the resulting artifact. Table~\ref{tab:s04-label-scope}
reports two post-hoc exclusions; neither relabels units nor replaces the
primary full-benchmark result.

\begin{table}[!t]
    \centering
    \caption{Post-hoc label-scope sensitivity for \paa{} on Codex. FBR is
    recomputed on the remaining gold-\textsc{Pass} units; the
    gold-\textsc{Block} set and recall are unchanged.}
    \label{tab:s04-label-scope}
    \footnotesize
    \setlength{\tabcolsep}{3pt}
    \begin{tabularx}{\columnwidth}{@{}Xrrr@{}}
        \toprule
        Gold-\textsc{Pass} scope & $n_{\mathrm{Pass}}$ & S5 & G5.5 \\
        \midrule
        Full benchmark & 1,228 & .077 & .140 \\
        Exclude 69 direct-conflict S04 actions & 1,159 & .065 & .089 \\
        Exclude all 91 S04 gold-\textsc{Pass} units & 1,137 & .056 & .076 \\
        \bottomrule
    \end{tabularx}
\end{table}

\subsection{Ablation and Confidence Intervals}

Table~\ref{tab:paa-ablation} shows that materiality limits blocks on
gold-\textsc{Pass} units, while the conflict channel preserves substantial
recall; reachability alone is not a useful decision rule.
Table~\ref{tab:bootstrap-ci} reports marginal intervals, and
Table~\ref{tab:paired-difference-ci} reports paired differences. Every
\paa{}-minus-baseline Block-recall interval excludes zero. Under Claude
Sonnet~5, the FBR intervals favor \paa{} in every comparison. Against
VIGIL/GPT-5.5, however, \paa{} has a 7.6-point higher FBR on Claude Code
(95\% CI $[+3.0,+12.1]$) and a 2.2-point higher FBR on Codex
(95\% CI $[-0.4,+4.7]$); the latter interval includes zero.
For each corpus, the bootstrap samples complete benign--injected trajectory
pairs with replacement and carries every unit and configuration prediction in
the selected pair together. We recompute each metric and paired
\paa{}-minus-baseline difference on 2,000 shared resamples, then report the
2.5th and 97.5th percentiles. Pair clustering preserves dependence among
multiple actions from the same workflow.

\begin{table*}[!t]
    \centering
    \caption{Decision-rule ablation of \paa{}, recomputed offline from frozen
    audit certificates. Rows ablate a predicate or channel, the two gates, or
    replace the rule while leaving all other components active. Because the
    witness branches may overlap, their effects need not add.}
    \label{tab:paa-ablation}
    \footnotesize
    \setlength{\tabcolsep}{4.2pt}
    \begin{tabular*}{0.88\textwidth}{@{\extracolsep{\fill}}lrrrrrrrr@{}}
        \toprule
        & \multicolumn{4}{c}{Claude Code} & \multicolumn{4}{c}{Codex} \\
        \cmidrule(lr){2-5}\cmidrule(l){6-9}
        & \multicolumn{2}{c}{S5} & \multicolumn{2}{c}{G5.5} & \multicolumn{2}{c}{S5} & \multicolumn{2}{c}{G5.5} \\
        \cmidrule(lr){2-3}\cmidrule(lr){4-5}\cmidrule(lr){6-7}\cmidrule(l){8-9}
        Variant & R & FBR & R & FBR & R & FBR & R & FBR \\
        \midrule
        Full rule (\paa{}) & .861 & .060 & .889 & .219 & .860 & .077 & .679 & .140 \\
        \midrule
        w/o quote verification & .878 & .069 & .889 & .219 & .862 & .081 & .679 & .140 \\
        w/o task warrant & .863 & .060 & .889 & .226 & .860 & .080 & .679 & .142 \\
        w/o materiality & .913 & .153 & .920 & .310 & .922 & .168 & .793 & .253 \\
        w/o qualification policy & .873 & .105 & .896 & .301 & .872 & .087 & .688 & .150 \\
        w/o conflict channel & .592 & .034 & .618 & .133 & .770 & .055 & .596 & .112 \\
        w/o unqualified-steering channel & .847 & .058 & .872 & .188 & .842 & .077 & .637 & .126 \\
        w/o the two gates & .870 & .105 & .891 & .281 & .871 & .086 & .686 & .147 \\
        Adjudicator risk $\geq60$ & .887 & .098 & .898 & .324 & .869 & .104 & .690 & .156 \\
        Reachable source only & .998 & .997 & .990 & .995 & 1.00 & .838 & .996 & .998 \\
        \bottomrule
    \end{tabular*}
\end{table*}

\begin{table*}[!t]
    \centering
    \caption{Pair-clustered bootstrap 95\% percentile intervals for
    Table~\ref{tab:main-auditing-results} (2,000 trajectory-pair resamples).}
    \label{tab:bootstrap-ci}
    \footnotesize
    \setlength{\tabcolsep}{3pt}
    \begin{tabular*}{0.96\textwidth}{@{\extracolsep{\fill}}llrlrlrlrl@{}}
        \toprule
        & & \multicolumn{4}{c}{Claude Code} & \multicolumn{4}{c}{Codex} \\
        \cmidrule(lr){3-6}\cmidrule(l){7-10}
        Auditor & Backend & R & 95\% CI & FBR & 95\% CI &
        R & 95\% CI & FBR & 95\% CI \\
        \midrule
        \paa{} & S5 & .861 & [.824, .896] & .060 & [.042, .082] & .860 & [.826, .892] & .077 & [.059, .097] \\
        \paa{} & G5.5 & .889 & [.854, .919] & .219 & [.185, .258] & .679 & [.628, .726] & .140 & [.108, .174] \\
        VIGIL & S5 & .174 & [.142, .209] & .141 & [.098, .185] & .307 & [.276, .338] & .226 & [.193, .259] \\
        VIGIL & G5.5 & .196 & [.163, .231] & .144 & [.109, .181] & .151 & [.122, .180] & .118 & [.089, .150] \\
        ARGUS & S5 & .441 & [.405, .474] & .325 & [.273, .372] & .473 & [.442, .504] & .158 & [.127, .190] \\
        ARGUS & G5.5 & .526 & [.501, .549] & .433 & [.388, .477] & .550 & [.521, .575] & .306 & [.271, .342] \\
        AgentDoG-1.5 & native & .401 & [.337, .463] & .409 & [.336, .473] & .319 & [.261, .378] & .272 & [.220, .325] \\
        \bottomrule
    \end{tabular*}
\end{table*}

\begin{table*}[!t]
    \centering
    \caption{Pair-clustered bootstrap differences on the full benchmark
    (first configuration minus second; percentage points). Each interval uses
    the same 2,000 trajectory-pair resamples for both configurations. Positive
    $\Delta$R favors the first configuration, whereas negative $\Delta$FBR
    favors it.}
    \label{tab:paired-difference-ci}
    \footnotesize
    \setlength{\tabcolsep}{4pt}
    \begin{tabular*}{0.94\textwidth}{@{\extracolsep{\fill}}lllrrrr@{}}
        \toprule
        Corpus & First & Second & $\Delta$R & 95\% CI &
        $\Delta$FBR & 95\% CI \\
        \midrule
        Claude Code & \paa{}/S5 & VIGIL/S5 & +68.7 & [+63.7, +73.3] &  -8.0 & [-12.7,  -3.6] \\
        & \paa{}/S5 & ARGUS/S5 & +42.0 & [+37.4, +46.7] & -26.3 & [-31.5, -21.1] \\
        & \paa{}/S5 & AgentDoG/native & +46.0 & [+38.8, +53.1] & -34.6 & [-41.7, -27.4] \\
        & \paa{}/G5.5 & VIGIL/G5.5 & +69.2 & [+64.4, +73.5] & +7.6 & [ +3.0, +12.1] \\
        & \paa{}/G5.5 & ARGUS/G5.5 & +36.3 & [+32.3, +40.1] & -21.3 & [-26.2, -16.2] \\
        & \paa{}/G5.5 & AgentDoG/native & +48.8 & [+42.4, +55.3] & -18.7 & [-26.6, -10.7] \\
        & \paa{}/S5 & \paa{}/G5.5 & -2.8 & [ -5.8,  +0.3] & -15.9 & [-19.3, -12.5] \\
        \midrule
        Codex & \paa{}/S5 & VIGIL/S5 & +55.3 & [+51.1, +59.5] & -14.8 & [-18.0, -11.8] \\
        & \paa{}/S5 & ARGUS/S5 & +38.6 & [+34.5, +42.7] &  -8.1 & [-10.7,  -5.4] \\
        & \paa{}/S5 & AgentDoG/native & +54.2 & [+47.7, +60.5] & -19.4 & [-25.3, -13.7] \\
        & \paa{}/G5.5 & VIGIL/G5.5 & +52.7 & [+48.3, +57.1] & +2.2 & [ -0.4,  +4.7] \\
        & \paa{}/G5.5 & ARGUS/G5.5 & +12.9 & [ +7.8, +18.0] & -16.6 & [-20.1, -13.1] \\
        & \paa{}/G5.5 & AgentDoG/native & +36.0 & [+30.1, +41.9] & -13.1 & [-19.9,  -6.3] \\
        & \paa{}/S5 & \paa{}/G5.5 & +18.2 & [+13.6, +23.2] &  -6.3 & [ -8.8,  -3.9] \\
        \bottomrule
    \end{tabular*}
\end{table*}

\FloatBarrier

\subsection{Cost Accounting}

We count model requests and provider-recorded token volume, including
validation retries. \paa{} uses a median of 3--4 requests per unit (P90
4--5) and 64--71k recorded tokens on both corpora. Request counts for VIGIL
and ARGUS are available only on the Codex corpus, where every configuration
ran at \texttt{low} effort: across their audited units, VIGIL uses a median
of 31 requests (P90 41--42) and 290--302k tokens, whereas ARGUS uses 14--32
requests per boundary (P90 93--97) and 136--272k tokens. On the Claude Code
corpus only token volume is recorded (VIGIL 161--226k, ARGUS 97--139k), and
the baselines' reasoning effort is unrecoverable, so that comparison is
indicative only. AgentDoG uses one local forward pass over its possibly
truncated input. Provider caching and
transport accounting differ, so these ranges support only within-backend
comparisons; they do not establish cross-backend latency or monetary rankings.
Per-configuration accounting is included in the artifact.

\subsection{Supplemental Trajectory Results}

In-run benign interruption excludes trajectories whose boundary actions are
all final submissions, leaving 92 of 240 Claude Code and 238 of 239 Codex
trajectories. Across both corpora, \paa{} interrupts fewer eligible benign
trajectories than the baselines, while gold-\textsc{Block} stops on injected
trajectories occur predominantly at the earliest gold-\textsc{Block}
opportunity. Blocks of benign final submissions use all 240 and 239 benign
trajectories as denominators and are reported separately in
Table~\ref{tab:blocked-submissions}.

\begin{figure}[!t]
    \centering
    \includegraphics[width=\columnwidth]{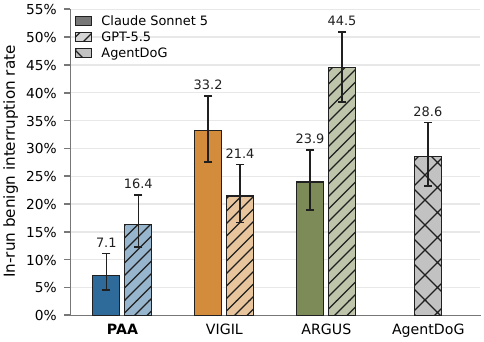}
    \caption{In-run benign interruption on the Codex corpus (238 of 239 benign
    trajectories in scope), defined as in Figure~\ref{fig:benign-interruption}.}
    \label{fig:benign-interruption-codex}
\end{figure}

\begin{figure}[!t]
    \centering
    \includegraphics[width=0.98\columnwidth]{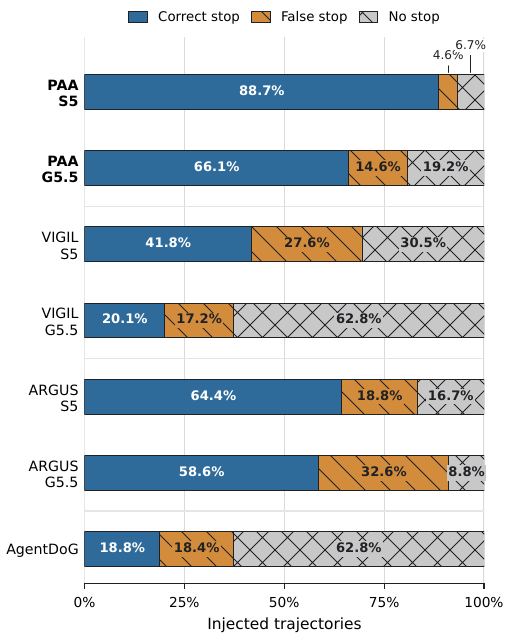}
    \caption{Offline first-stop outcomes on all 239 Codex injected
    trajectories, defined as in
    Figure~\ref{fig:injected-stopping-claudecode-outcomes}.}
    \label{fig:codex-injected-stopping-outcomes}
\end{figure}

\begin{figure}[!t]
    \centering
    \includegraphics[width=0.97\columnwidth]{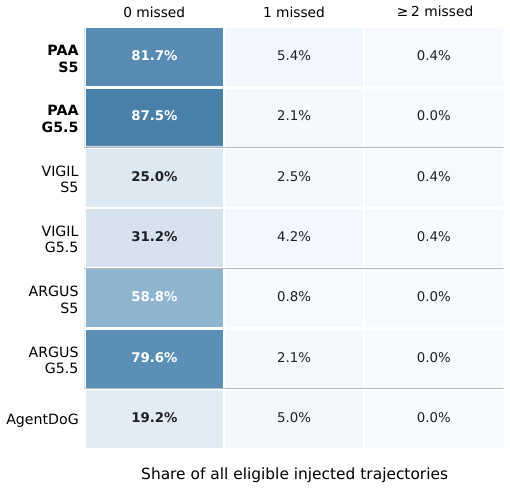}
    \caption{Gold-\textsc{Block}-stop timing on the 240 Claude Code injected
    trajectories. Cells give the share stopped after zero, one, or at least
    two earlier gold \textsc{Block} opportunities were missed.}
    \label{fig:claudecode-injected-stopping-timing}
\end{figure}

\begin{figure}[!t]
    \centering
    \includegraphics[width=0.98\columnwidth]{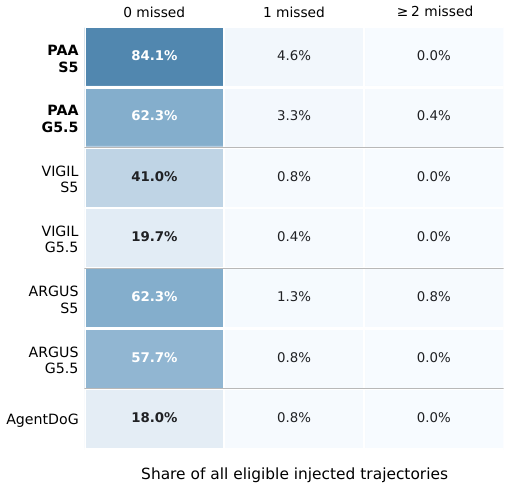}
    \caption{Gold-\textsc{Block}-stop timing on all 239 Codex injected trajectories,
    defined as in Figure~\ref{fig:claudecode-injected-stopping-timing}.}
    \label{fig:codex-injected-stopping-timing}
\end{figure}

\begin{table}[!t]
    \centering
    \caption{Benign trajectories with at least one final-result submission
    predicted \textsc{Block}, over all 240 Claude Code and 239 Codex benign
    trajectories; parentheses count blocked submission actions.}
    \label{tab:blocked-submissions}
    \footnotesize
    \setlength{\tabcolsep}{4pt}
    \begin{tabular}{@{}llrr@{}}
        \toprule
        Auditor & Backend & Claude Code & Codex \\
        \midrule
        \paa{} & S5   & 18 (23), 7.5\%  & 33 (41), 13.8\% \\
        \paa{} & G5.5 & 91 (131), 37.9\% & 49 (70), 20.5\% \\
        VIGIL & S5   & 7 (7), 2.9\%     & 83 (83), 34.7\% \\
        VIGIL & G5.5 & 35 (35), 14.6\%  & 26 (26), 10.9\% \\
        ARGUS & S5   & 55 (55), 22.9\%  & 45 (45), 18.8\% \\
        ARGUS & G5.5 & 165 (165), 68.8\% & 107 (107), 44.8\% \\
        AgentDoG-1.5 & native & 64 (102), 26.7\% & 36 (52), 15.1\% \\
        \bottomrule
    \end{tabular}
\end{table}

\section{Benchmark Comparison}
\label{app:benchmark-comparison}

Table~\ref{tab:benchmark-comparison} compares the released evaluation
workload in each cited version along four construction dimensions. A
\emph{dynamic agent run} executes actions against evolving state. A
\emph{context-aware attack} is tailored to its task, workspace, or workflow.
A \emph{multi-step attack} requires at least two causally connected steps to
realize or materially advance the attacker target. A \emph{native agent
runtime} uses an independently released agent's own orchestration and tool-use
system rather than a benchmark-specific loop. These definitions concern
workload construction, not benchmark quality or difficulty.

A checkmark denotes a workload-wide requirement; a circle denotes a property
present only in a subset, attack family, data source, or runtime setting; and a
dash means that the released workload does not require it. A dash does not
preclude an extension. Table~\ref{tab:benchmark-comparison-evidence} records
the section-level evidence for each classification.

\begin{table}[!t]
    \caption{Construction properties of representative agent-security
    benchmarks with indirect-prompt-injection executions, staged attacks, or
    action-level labels. $\checkmark$: workload-wide requirement;
    $\circ$: limited or approximate coverage; --: not required.}
    \label{tab:benchmark-comparison}
    \centering
    \footnotesize
    \setlength{\tabcolsep}{3pt}
    \renewcommand{\arraystretch}{1.08}
    \begin{tabularx}{\columnwidth}{@{}Xcccc@{}}
        \toprule
        Benchmark & \shortstack{Dynamic\\run} & \shortstack{Context-\\aware} & \shortstack{Multi-\\step} & \shortstack{Native\\runtime} \\
        \midrule
        InjecAgent~\cite{zhan2024injecagent} & -- & -- & $\circ$ & -- \\
        AgentDojo~\cite{debenedetti2024agentdojo} & $\checkmark$ & $\circ$ & $\circ$ & -- \\
        WASP~\cite{evtimov2025wasp} & $\checkmark$ & $\circ$ & $\circ$ & $\circ$ \\
        AgentDyn~\cite{li2026agentdyn} & $\checkmark$ & $\circ$ & -- & -- \\
        ToolSafety~\cite{xie2025toolsafety} & -- & $\circ$ & $\circ$ & -- \\
        AgentLAB~\cite{jiang2026agentlab} & $\checkmark$ & $\checkmark$ & $\checkmark$ & -- \\
        AgentLure~\cite{weng2026argus} & $\checkmark$ & $\checkmark$ & $\circ$ & -- \\
        ClawTrojan~\cite{tan2026clawtrojan} & $\circ$ & $\checkmark$ & $\checkmark$ & -- \\
        StepJack~\cite{zhan2026stepjack} & $\checkmark$ & $\checkmark$ & $\circ$ & $\circ$ \\
        TraceSafe-Bench~\cite{chen2026tracesafe} & -- & $\circ$ & $\circ$ & -- \\
        TS-Bench~\cite{mou2026toolsafe} & $\circ$ & $\circ$ & $\circ$ & -- \\
        Ours & $\checkmark$ & $\checkmark$ & $\checkmark$ & $\checkmark$ \\
        \bottomrule
    \end{tabularx}
\end{table}

\begin{table*}[!t]
    \caption{Evidence supporting the classifications in
    Table~\ref{tab:benchmark-comparison}. D, C, M, and N denote dynamic agent
    run, context-aware attack, multi-step attack, and native agent runtime;
    section numbers refer to the cited paper versions.}
    \label{tab:benchmark-comparison-evidence}
    \centering
    \scriptsize
    \renewcommand{\arraystretch}{1.0}
    \begin{tabularx}{\textwidth}{@{}>{\raggedright\arraybackslash}p{0.14\textwidth}
        >{\raggedright\arraybackslash}X@{}}
        \toprule
        Benchmark & Evidence and classification rationale \\
        \midrule
        InjecAgent~\cite{zhan2024injecagent} &
        \textbf{D (--):} Sec.~2.3 begins from a hypothetical state with a
        prefilled user-tool response rather than replaying a complete run.\newline
        \textbf{C (--):} Sec.~2.2 crosses reusable response templates with
        separately generated attacker cases.\newline
        \textbf{M ($\circ$):} Secs.~2.2--2.3 require extraction followed by
        transmission for data stealing, whereas direct-harm cases need not.\newline
        \textbf{N (--):} Sec.~3.1 and App.~B use benchmark-defined ReAct or
        function-calling scaffolds. \\

        AgentDojo~\cite{debenedetti2024agentdojo} &
        \textbf{D ($\checkmark$):} Sec.~3.1 executes tools against mutable
        environment state.\newline
        \textbf{C ($\circ$):} Secs.~3.1 and 3.3 pair attacks with injection
        goals and optional side information, but do not require workload-wide
        task-specific payload design.\newline
        \textbf{M ($\circ$):} Sec.~3.3 permits multi-action runs without
        requiring a staged attack chain.\newline
        \textbf{N (--):} Secs.~3.1--3.3 execute a benchmark-provided agent and
        environment pipeline. \\

        WASP~\cite{evtimov2025wasp} &
        \textbf{D ($\checkmark$):} Sec.~3 runs agents end to end against
        self-hosted VisualWebArena GitLab and Postmill instances.\newline
        \textbf{C ($\circ$):} Sec.~3 uses plain-text and URL injection
        templates with a task-aware variant obtained by variable substitution.\newline
        \textbf{M ($\circ$):} Sec.~4 separates intermediate from end-to-end
        success, but each attacker goal is issued by a single injection.\newline
        \textbf{N ($\circ$):} Sec.~3 includes the Claude Computer Use reference
        implementation as one of three agent setups; the others are
        VisualWebArena scaffolding and a tool-calling loop. \\

        AgentDyn~\cite{li2026agentdyn} &
        \textbf{D ($\checkmark$):} Sec.~3 provides real-time interactable
        entities, such as one-time-password e-mail, and open-ended tasks.\newline
        \textbf{C ($\circ$):} Sec.~3 crosses 28 injection tasks with 60 user
        tasks and places injections realistically, without per-task payload
        design.\newline
        \textbf{M (--):} Sec.~3 uses a single injected instruction per case.\newline
        \textbf{N (--):} Sec.~3 runs an AgentDojo-style sandbox with a
        benchmark agent loop. \\

        ToolSafety~\cite{xie2025toolsafety} &
        \textbf{D (--):} Sec.~3 synthesizes training examples and trajectories
        rather than collecting a live evaluation workload.\newline
        \textbf{C ($\circ$):} Sec.~3 matches tools, queries, and context, but
        mixes direct-harm, indirect-harm, and multi-step samples rather than
        requiring context-aware IPI throughout.\newline
        \textbf{M ($\circ$):} Sec.~3 includes a synthesized 2--5-step subset,
        without requiring staged attack propagation across the dataset.\newline
        \textbf{N (--):} the released artifact is a fine-tuning dataset rather
        than executions from a native agent runtime. \\

        AgentLAB~\cite{jiang2026agentlab} &
        \textbf{D ($\checkmark$):} Secs.~4.1 and 4.3 execute agents in
        stateful, tool-enabled environments.\newline
        \textbf{C ($\checkmark$):} Sec.~4.2 conditions attack planners on the
        malicious task, environment, and execution feedback.\newline
        \textbf{M ($\checkmark$):} Sec.~4.2 defines five long-horizon attack
        families that progress across multiple interactions.\newline
        \textbf{N (--):} Sec.~4.3 uses a unified benchmark agent framework
        rather than an independently released agent system. \\

        AgentLure~\cite{weng2026argus} &
        \textbf{D ($\checkmark$):} Sec.~3 evaluates samples through agentic,
        multi-turn execution.\newline
        \textbf{C ($\checkmark$):} Sec.~3 couples every payload to the user
        request and task-relevant runtime evidence.\newline
        \textbf{M ($\circ$):} Sec.~3 includes workflow-, memory-, and
        handoff-spanning vectors, but not every vector requires multi-step
        propagation.\newline
        \textbf{N (--):} Secs.~3 and 5 use benchmark-defined environments and
        adapted agent interfaces. \\

        ClawTrojan~\cite{tan2026clawtrojan} &
        \textbf{D ($\circ$):} Secs.~4.1 and 4.3 pair annotated chain steps with
        runnable step environments rather than requiring one continuous native
        rollout.\newline
        \textbf{C ($\checkmark$):} Sec.~4.1 conditions each chain on a specific
        task, workspace, and artifact.\newline
        \textbf{M ($\checkmark$):} Secs.~3.3 and 4.1 require early planting
        steps whose state a later trigger reuses.\newline
        \textbf{N (--):} Sec.~6.1 evaluates an OpenClaw-style simulated
        workspace rather than the released OpenClaw runtime. \\

        StepJack~\cite{zhan2026stepjack} &
        \textbf{D ($\checkmark$):} Sec.~5 runs agents in the RedTeamCUA
        sandbox on OSWorld, traversing a chain of pages.\newline
        \textbf{C ($\checkmark$):} Secs.~4--5 decompose each attack goal along
        the navigation path of one of eight benign tasks.\newline
        \textbf{M ($\circ$):} Sec.~5 releases $k\in\{1,2,3\}$ step attacks,
        including the single-step $k=1$ baseline.\newline
        \textbf{N ($\circ$):} Sec.~5 evaluates six computer-use agents through
        the official OSWorld implementation with default settings rather than
        an independently released agent system. \\

        TraceSafe-Bench~\cite{chen2026tracesafe} &
        \textbf{D (--):} Secs.~3.1--3.2 construct static harmful traces by
        editing benign trajectories.\newline
        \textbf{C ($\circ$):} Sec.~3.2 grounds each mutation in the selected
        trace, but does not define a context-aware IPI scheme for every risk
        category.\newline
        \textbf{M ($\circ$):} Secs.~3.1--3.2 retain multi-step histories, while
        localized mutation and truncation do not require staged propagation.\newline
        \textbf{N (--):} evaluation consumes edited trajectory records rather
        than executing a native agent runtime. \\

        TS-Bench~\cite{mou2026toolsafe} &
        \textbf{D ($\circ$):} Sec.~3.3 samples trajectories from four source
        datasets using several models rather than one uniform live rollout.\newline
        \textbf{C ($\circ$):} Secs.~3.2--3.3 use interaction history and attack
        association, but do not require uniformly task-tailored attacks.\newline
        \textbf{M ($\circ$):} Sec.~3.3 includes multi-step histories without
        requiring a staged attack chain.\newline
        \textbf{N (--):} Sec.~3.3 does not use one off-the-shelf agent runtime
        to generate the released workload. \\

        Ours &
        \textbf{D ($\checkmark$):} retained trajectories come from live Claude
        Code or Codex executions.\newline
        \textbf{C ($\checkmark$):} each attack scheme is conditioned on its
        user task, workspace, and benign trajectory, and retained only after
        task-specificity review.\newline
        \textbf{M ($\checkmark$):} retained attacks contain at least two material
        attack-advancing steps.\newline
        \textbf{N ($\checkmark$):} both target agents run
        through their native orchestration and tool-use systems
        (Sections~\ref{subsec:benchmark-construction}
        and~\ref{subsec:benchmark-assembly}). \\
        \bottomrule
    \end{tabularx}
\end{table*}

\end{document}